\documentclass[preprint2]{aastex62}

\usepackage{amsmath, amsthm, amssymb, amsfonts}

\newcommand{\Rp}{R_p}

\newcommand{\Rs}{R_{\star}}
\newcommand{\aRs}{a/R_{\star}}

\newcommand{\RpRs}{\Rp/\Rs}

\newcommand{\RE}{R_{\Earth}}
\newcommand{\ME}{M_{\Earth}}

\newcommand{\um}{\mu\textnormal{m}}
\newcommand{\Teff}{T_{\textnormal{eff}}}
\newcommand{\FeH}{[{\rm{Fe}/\rm{H}}]}
\newcommand{\logg}{\log_{10} g}
\newcommand{\XHtwoO}{X_{\rm{H}_2\rm{O}}}

\received{July 5, 2022}
\revised{October 30, 2022}
\accepted{November 22, 2022}
\submitjournal{AAS Journals}

\shorttitle{Transmission spectroscopy for the temperate sub-Neptune TOI-270\,d}

\shortauthors{Mikal-Evans et al.}

\begin{document}

\title{Hubble Space Telescope transmission spectroscopy for the temperate sub-Neptune TOI-270d: a possible hydrogen-rich atmosphere containing water vapour}

\correspondingauthor{Thomas Mikal-Evans}
\email{tmevans@mpia.de}

\author[0000-0001-5442-1300]{Thomas Mikal-Evans}
\affiliation{Max Planck Institute for Astronomy, K\"{o}nigstuhl 17, D-69117 Heidelberg, Germany}

\author[0000-0002-4869-000X]{Nikku Madhusudhan}
\affiliation{Institute of Astronomy, University of Cambridge, Madingley Road, Cambridge CB3 0HA, UK}

\author{Jason Dittmann}
\affiliation{Department of Astronomy, University of Florida, 211 Bryant Space Science Center, Gainesville, FL 32611, USA}

\author[0000-0002-3164-9086]{Maximilian N. G\"unther}
\thanks{ESA Research Fellow}
\affiliation{European Space Agency (ESA), European Space Research and Technology Centre (ESTEC), Keplerlaan 1, 2201 AZ Noordwijk, The Netherlands}

\author[0000-0003-0156-4564]{Luis Welbanks}
\thanks{NHFP Sagan Fellow}
\affiliation{School of Earth \& Space Exploration, Arizona State University, Tempe, AZ, 85257, USA}

\author{Vincent Van Eylen}
\affiliation{Mullard Space Science Laboratory, University College London, Holmbury St Mary, Dorking, Surrey RH5 6NT, UK}

\author{Ian J.\ M.\ Crossfield}
\affiliation{Department of Physics and Astronomy, University of
  Kansas, Lawrence, KS, USA}

\author[0000-0002-6939-9211]{Tansu Daylan}
\thanks{LSSTC Catalyst Fellow}
\affiliation{Department of Astrophysical Sciences, Princeton University, 4 Ivy Lane, Princeton, NJ 08544}

\author{Laura Kreidberg}
\affiliation{Max Planck Institute for Astronomy, K\"{o}nigstuhl 17, D-69117 Heidelberg, Germany}


\begin{abstract}

TOI-270\,d is a temperate sub-Neptune discovered by the Transiting Exoplanet Survey Satellite (TESS) around a bright ($J=9.1$\,mag) M3V host star. With an approximate radius of $2\,\RE$ and equilibrium temperature of 350\,K, TOI-270\,d is one of the most promising small exoplanets for atmospheric characterisation using transit spectroscopy. Here we present a primary transit observation of TOI-270\,d made with the Hubble Space Telescope (HST) Wide Field Camera 3 (WFC3) spectrograph across the 1.126-1.644\,$\um$ wavelength range, and a 95\% credible upper limit of $8.2 \times 10^{-14}$\,erg\,s$^{-1}$\,cm$^{-2}$\,\AA$^{-1}$\,arcsec$^{-2}$ for the stellar Ly$\alpha$ emission obtained using the Space Telescope Imaging Spectrograph (STIS). The transmission spectrum derived from the TESS and WFC3 data provides evidence for molecular absorption by a hydrogen-rich atmosphere at ${4\sigma}$ significance relative to a featureless spectrum. The strongest evidence for any individual absorber is obtained for H$_2$O, which is favoured at 3$\sigma$ significance. When retrieving on the WFC3 data alone and allowing for the possibility of a heterogeneous stellar brightness profile, the detection significance of H$_2$O is reduced to 2.8$\sigma$. Further observations are therefore required to robustly determine the atmospheric composition of TOI-270\,d and assess the impact of stellar heterogeneity. If confirmed, our findings would make TOI-270\,d one of the smallest and coolest exoplanets to date with detected atmospheric spectral features.

\end{abstract}

\section{Introduction} \label{sec:intro}

Sub-Neptunes are the population of planets with radii smaller than that of Neptune and larger than approximately $1.8\,\RE$. The latter corresponds to the location of the ``radius valley'' uncovered by the NASA \textit{Kepler} survey, which divides small planets between those with volatile-poor and volatile-rich compositions \citep{2017AJ....154..109F,2018MNRAS.479.4786V,2022AJ....163..179P}. Measured densities for the volatile-rich sub-Neptunes can be explained by a variety of compositions with varying core-mass fractions and envelopes of volatiles such as hydrogen and water \citep{2009A&A...493..671F,2010ApJ...716.1208R}. The measurement of sub-Neptune atmospheric spectra can help break this degeneracy by providing a direct probe of the composition of the uppermost layers of the outer envelope.

The observing strategy employed by the NASA Transiting Exoplanet Survey Satellite (TESS) \citep{2015JATIS...1a4003R} since 2018 is well suited for discovering short-period planets that subsequently make excellent targets for detailed atmospheric studies \citep{2015ApJ...809...77S, 2018ApJS..239....2B, 2021ApJS..254...39G, 2022AJ....163..290K}. Most significantly, TESS has observed around 90\% of the sky, meaning that a substantial fraction of all nearby bright stars have now been monitored for planetary transits with complete phase coverage out to orbital periods of approximately 10 days. This includes the majority of nearby M dwarfs, around which smaller planets such as sub-Neptunes can be more readily detected compared to those orbiting earlier spectral type hosts, thanks to the relatively deep transit signals that they produce. As of October 2022, the NASA Exoplanet Archive\footnote{https://exoplanetarchive.ipac.caltech.edu} reports that there have been seventeen transiting exoplanets with radii falling in the 1.8-4$\,\RE$ sub-Neptune range, validated around bright ($J<10$\,mag) M dwarfs. Of these, TESS has discovered twelve, namely: AU~Mic\,b and AU~Mic\,c \citep{2020Natur.582..497P,2021A&A...649A.177M}; TOI-1634\,b \citep{2021AJ....162...79C,2021AJ....162..161H}; TOI-270\,c and TOI-270\,d \citep{Guenther2019}; TOI-776\,b and TOI-776\,c \citep{2021A&A...645A..41L}; TOI-1201\,b \citep{2021A&A...656A.124K}; TOI-1231\,b \citep{2021AJ....162...87B}; TOI-1266\,b \citep{2020A&A...642A..49D,2020AJ....160..259S}; LTT3780\,c \citep{2020AJ....160....3C,2020A&A...642A.173N}; and GJ~3090\,b \citep{2022A&A...665A..91A}. The other five planets that have been validated to date around bright M dwarfs and with radii between approximately 1.8-4$\,\RE$ are: GJ~436\,b \citep{2007A&A...472L..13G}; GJ~1214\,b \citep{2009Natur.462..891C}; K2-3\,b and K2-3\,c \citep{2015ApJ...804...10C}; and K2-18\,b \citep{2015ApJ...809...25M}. Meanwhile, there are currently 57 additional sub-Neptune candidates that have been identified by TESS around bright M dwarfs and that are listed as high priority targets for follow-up confirmation.\footnote{https://exofop.ipac.caltech.edu/tess}

This paper presents transmission spectroscopy measurements for the TESS-discovered sub-Neptune TOI-270\,d/L\,231-32\,d \citep{Guenther2019,VanEylen2021,Kaye2022}, made using the \textit{Hubble Space Telescope} Wide Field Camera 3 (HST WFC3). TOI-270\,d has a radius of $2.00 \pm 0.05\,\RE$ measured in the TESS red-optical passband and a mass of $4.20 \pm 0.16\,\ME$, corresponding to a bulk density of $2.90 \pm 0.24$\,g\,cm$^{-3}$ \citep{Kaye2022}. It orbits an M3V host star at a distance of 0.07\,AU with an orbital period of $11.4$ days. This places TOI-270\,d just inside the inner edge of the habitable zone, where it has an equilibrium temperature of around 350\,K assuming a Bond albedo of 30\% and uniform day-night heat redistribution.\footnote{Earth, Jupiter, Saturn, Uranus, and Neptune all have Bond albedos close to 30\%.} There are two other transiting planets known to orbit the same host star \citep{Guenther2019}: TOI-270\,b, a $1.3\,\RE$ super-Earth on a 0.03\,AU orbit; and TOI-270\,c, a $2.3\,\RE$ sub-Neptune on a 0.05\,AU orbit. The TOI-270 host star appears to be quiescent, with TESS photometry showing no evidence for variability, either due to flares or brightness fluctuations caused by stellar heterogeneities (i.e.\ faculae and spots) rotating into and out of view \citep{Guenther2019}. Additional indicators --- such as H$\alpha$, the S-index, and a lack of radial velocity jitter (root-mean-square of $0.16 \pm 0.23$\,m\,s$^{-1}$) --- are also consistent with a low stellar activity level \citep{VanEylen2021}. To further characterise the host star, in this study we also present new constraints for the stellar Ly$\alpha$ emission obtained with the HST Space Telescope Imaging Spectrograph (STIS).

The paper is organised into the following sections. Observations are described in Sections \ref{sec:observations}. Analyses of the WFC3 and STIS data are presented in Sections \ref{sec:dataredWFC3} and \ref{sec:dataredSTIS}, respectively. Atmospheric retrieval analyses are presented in Section \ref{sec:retrieval}. Discussion and conclusions are given in Sections \ref{sec:discussion} and \ref{sec:conclusion}, respectively.

\section{Observations} \label{sec:observations}

A transit of TOI-270\,d was observed with HST on 2020 October 6 as part of program GO-15814 \citep{2019hst..prop15814M}. Data were acquired over three consecutive HST orbits, with the first and third orbits providing, respectively, pre-transit and post-transit reference levels for the stellar brightness. The second HST orbit coincided with the planetary transit, which has a duration of 2.117 hours \citep{Kaye2022}. 

The transit observation was made using the infrared channel of WFC3 with the G141 grism, covering the 1.126-1.644\,$\um$ wavelength range. During each exposure, the spectral trace was allowed to drift along the cross-dispersion axis of the detector using the round-trip spatial-scanning observing mode \citep{2013ApJ...774...95D,2014ApJ...791...55M}, with a scan rate of 0.111 arcsec s$^{-1}$. A $512 \times 512$ pixel subarray containing the target spectrum was read out for each exposure using the SPARS25 pattern with seven non-destructive reads per exposure (NSAMP=7), corresponding to individual exposure times of 138 seconds. With this setup, 15 exposures were taken during the first HST orbit following target acquisition, and 16 exposures were taken in both the second and third HST orbits. A single shorter exposure (NSAMP=3) was also taken at the end of the second and third HST orbits, due to anticipated overrun of the target visibility window (i.e.\ the target being lost from view before a full exposure could be completed). These latter shorter exposures were ultimately discarded from the subsequent analysis.

To characterise the Ly$\alpha$ emission of the host star, two additional observations were made with HST on 2020 January 29 and 31 as part of the same observing program, using HST STIS with the $52 \times 0.05$ arcsec slit and the G140M grating centered at 1,222\AA. For each STIS observation, data were collected for 1,960 seconds using TIME-TAG mode. Both exposures were timed to avoid transits of the three known planets in the TOI-270 system, as the goal was to calibrate the absolute emission level of the host star.

\section{WFC3 analysis: transmission spectroscopy} \label{sec:dataredWFC3}

The WFC3 transit data were reduced using the custom Python code described previously in \cite{2016ApJ...822L...4E} and since used in numerous other studies \citep[e.g.][]{2019MNRAS.488.2222M,2020MNRAS.496.1638M}. Individual exposures were provided by the WFC3 pipeline as FITS files with IMA suffixes, for which basic calibrations have already been applied, such as flat fielding, linearity correction, and bias subtraction. For a given calibrated IMA exposure, background counts were determined for each non-destructive read by taking the median value of a $20 \times 120$ pixel box away from the target spectrum. For each IMA frame, the differences between successive, background-subtracted, non-destructive reads were computed. These read-differences correspond to ``stripes'' of flux that span the cross-dispersion rows scanned between each read. The flux-weighted cross-dispersion centroids of these stripes were determined and all rows more than 50 pixels above and below the stripe centroids were set to zero to mask contaminating contributions, such as neighbouring stars and cosmic ray strikes. The masked stripes were then summed together to form reconstructed data frames. For each of these reconstructed data frames, the final flux-weighted cross-dispersion centroid was determined and 1D spectra were extracted by summing all pixels along the cross-dispersion axis within 120 pixels above and below the centroid row. The wavelength solution was obtained by cross-correlating the 1D spectrum extracted from the final exposure against a model stellar spectrum multiplied by the G141 throughput profile. The Python package \texttt{pysynphot} \citep{2013ascl.soft03023S} was used to obtain the model stellar spectrum: namely, a Kurucz 1993 model spectrum with properties close to those reported by \cite{Guenther2019} for the TOI-270 host star ($\Teff=3500$\,K, $\logg = 4.88$ cgs, $\FeH=-0.17$ dex).

\subsection{Broadband transit light curve} \label{sec:lcBroad}

A broadband light curve was generated by integrating the flux for each 1D spectrum across the $0.8$-$1.95\,\um$ wavelength range, to conservatively encompass the full G141 passband. The light curve was then fitted by simultaneously modelling the planetary transit signal and instrumental systematics. 

For the planetary transit signal, we used the \texttt{batman} software package \citep{2015PASP..127.1161K}. The planetary parameters allowed to vary in the fitting were: the planet-to-star radius ratio ($\RpRs$); the deviation of the transit mid-time ($\Delta T_c$) from the mid-time predicted by \cite{Kaye2022}; the normalised semimajor axis ($\aRs$); and the transit impact parameter ($b=a\cos i/\Rs$, where $i$ is the orbital inclination). Uniform priors were adopted for $\RpRs$ and $\Delta T_c$. The posterior distributions for $\aRs$ and $b$ obtained by \cite{Kaye2022} were adopted as Gaussian priors in the light curve fitting, namely: $\aRs = 41.744 \pm 0.527$ and $b = 0.23 \pm 0.08$. The orbital period was held fixed to $P=11.38194$ based on the value reported by \cite{Kaye2022}. 

For the stellar brightness profile, we assumed a quadratic limb darkening law. Both limb darkening coefficients were allowed to vary during the light curve fitting, but due to the incomplete phase coverage of the transit, tight Gaussian priors were adopted. The latter were obtained by providing the measured values and uncertainties for the stellar properties reported by \cite{VanEylen2021} as input to the PyLDTK software package \citep{Parviainen2015}, along with the G141 throughput profile. This resulted in Gaussian priors of the form $u_1 = 0.1611 \pm 0.0002$ and $u_2 = 0.1379 \pm 0.0008$.

\begin{table}
\begin{minipage}{\columnwidth}
  \centering
\scriptsize
\caption{TOI-270\,d system properties and broadband light curve fit results. Fixed properties have been taken from \cite{Guenther2019} and \cite{Kaye2022}.  \label{table:wfc3lcFitBroadRaw}}
\begin{tabular}{ccc}
\hline        Fixed &                                          Parameter &                               Value \\ \cline{2-3} 
             &                          $R_{\star}$ ($R_{\odot}$) &                             $0.380$ \\ 
             &                                            $P$ (d) &                          $11.38014$ \\ 
             &                                                $e$ &                                 $0$ \\ 
\hline         Free &                                          Parameter &                           Posterior \\ \cline{2-3} 
             &                                    $R_p/R_{\star}$ &     $0.05278_{-0.00118}^{+0.00095}$ \\ 
             &                                      $a/R_{\star}$ &             $41.80_{-0.59}^{+0.61}$ \\ 
             &                                                $b$ &           $0.233_{-0.085}^{+0.089}$ \\ 
             &                                 $\Delta T_1$ (min) &                     $-4_{-8}^{+10}$ \\ 
             &                                              $u_1$ &        $0.1611_{-0.0002}^{+0.0002}$ \\ 
             &                                              $u_2$ &        $0.1378_{-0.0009}^{+0.0009}$ \\ 
             &                                          $\beta$ &                 $3.0_{-0.4}^{+0.5}$ \\ 
\hline Derived &                                        Parameter &               Posterior \\ \cline{2-3} 
             &                          $(R_p/R_{\star})^2$ (ppm) &                $2786_{-124}^{+101}$ \\ 
             &                               $R_p$ ($R_{\Earth}$) &              $2.19_{-0.07}^{+0.07}$ \\ 
             &                                           $\cos i$ &        $0.0056_{-0.0020}^{+0.0021}$ \\ 
             &                                     $i$ ($^\circ$) &             $89.68_{-0.12}^{+0.12}$ \\ 
             &                $T_1$ (JD$_{{\textnormal{{UTC}}}}$) & $2459129.34868_{-0.00544}^{+0.00691}$ \\  \hline

\end{tabular}
\end{minipage}
\end{table}

We modelled the systematics using a Gaussian process (GP) model, with covariance described by the sum of two squared-exponential kernels that took time $t$ and the HST orbital phase $\varphi$ as input variables. For the GP mean function, the transit signal was multiplied by a deterministic systematics signal of the form $S_d( t, \varphi ) = (c_d+\gamma t ) \, D(t ,  \varphi  )$, where: $c_d$ is a normalisation constant for scan direction $d$ (i.e. forward or backward); $\gamma$ is the slope of a linear $t$-dependent baseline trend shared by the two scan directions; and $D$ is an analytic model for the electron charge-trapping systematics that affect the WFC3 detector \citep{2017AJ....153..243Z}. For the latter, the same double-exponential ramp function described in \cite{2022NatAs...6..471M} was adopted, which has five free parameters ($a_1$, $a_2$, $a_3$, $a_4$, $a_5$). To allow for additional high-frequency noise not captured by the systematics model, the Gaussian measurement uncertainties were allowed to vary in the fitting via a free parameter $\beta$ that rescaled the photon noise. For further details on the implementation of this systematics treatment, see \cite{2021AJ....161...18M}.

\begin{figure}
\centering  
\includegraphics[width=\columnwidth]{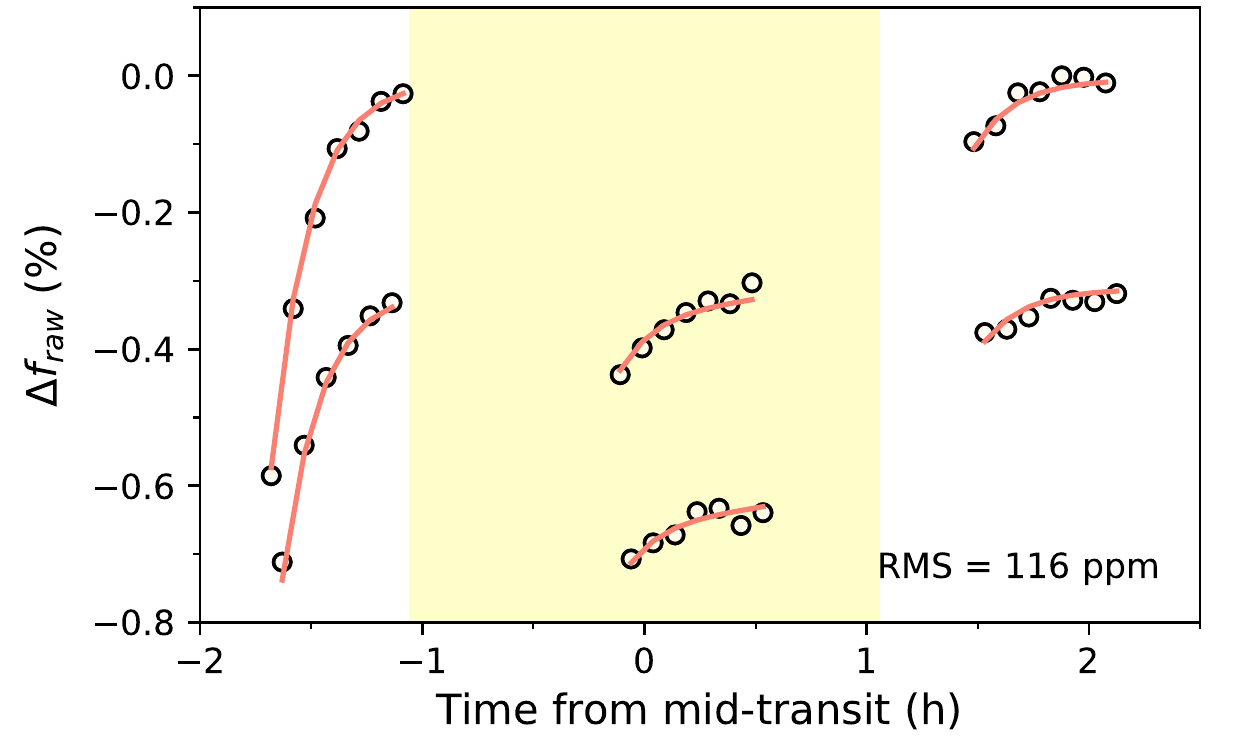}
\caption{Circles show the raw WFC3 broadband transit light curve for TOI-270\,d as the relative variation of received flux over time. There is an offset in the flux levels measured for the forward scan and backward scan exposures. Red lines show the maximum-likelihood light curve model. The root-mean-square (RMS) of the model fit residuals is printed in the lower right corner of the axis. Yellow shading indicates the time between transit ingress and egress.}
\label{fig:wfc3lcFitBroadRaw}
\end{figure}

\begin{figure}
\centering  
\includegraphics[width=\columnwidth]{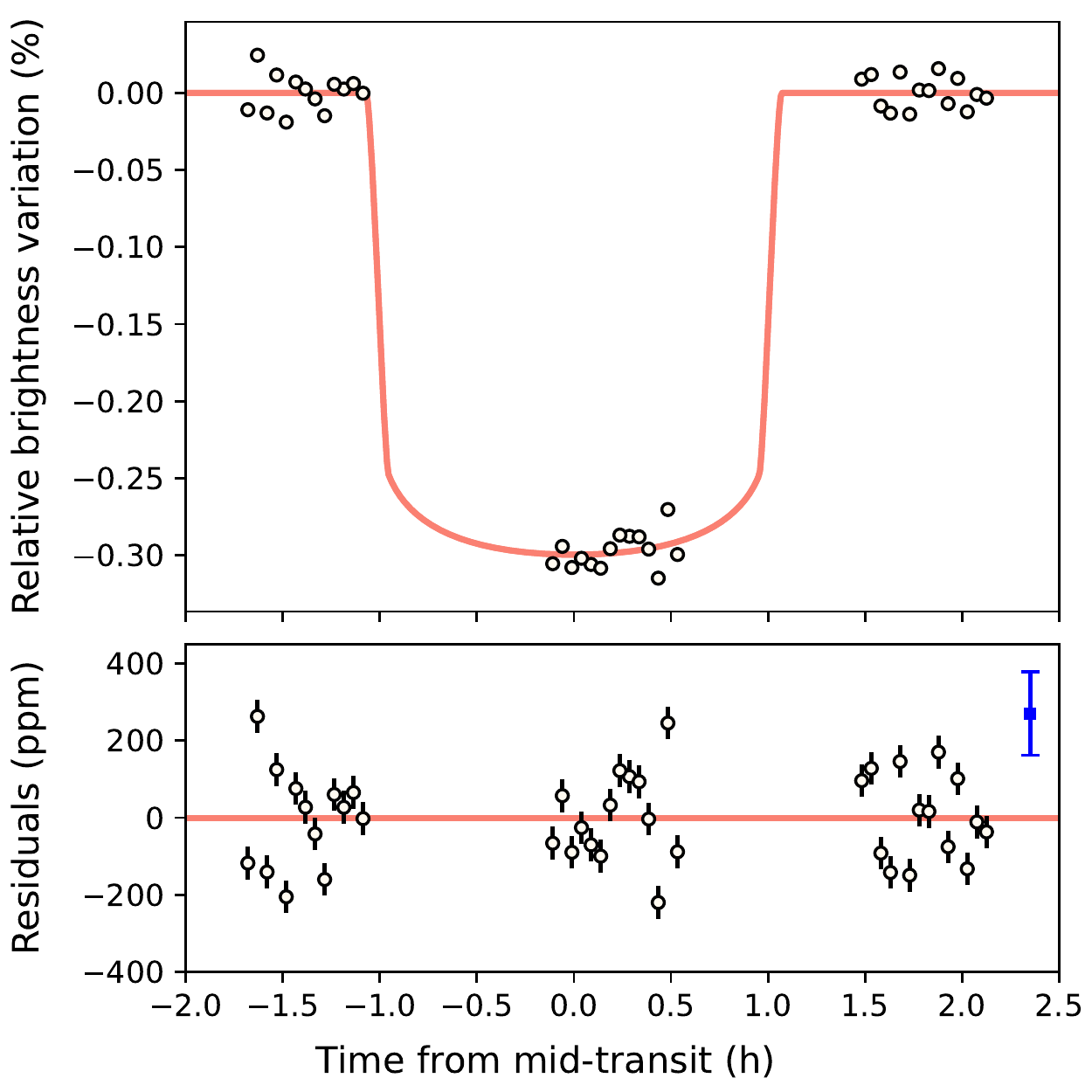}
\caption{(a) Circles show the measured broadband light curve after dividing through by the systematics component of the maximum-likelihood model. Measurement uncertainties are approximately the same size as the circles. Red line shows the transit component of the maximum-likelihood model. (b) Circles show residuals of the maximum-likelihood model with $1\sigma$ photon noise measurement uncertainties. Blue square with errorbar on the right of the axis indicates the size of the rescaled measurement uncertainties obtained from the model fit, i.e.\ the photon noise multiplied by the maximum likelihood $\beta$ value.}
\label{fig:wfc3lcFitBroadCorr}
\end{figure}

Uniform priors were adopted for all transit signal parameters, with the exception of the Gaussian priors described above for $\aRs$, $b$, $u_1$, and $u_2$. Marginalisation of the posterior distribution was performed using affine-invariant Markov chain Monte Carlo, as implemented by the \texttt{emcee} software package \citep{2013PASP..125..306F}, using 300 walkers and 1000 steps, with the first 500 steps discarded as burn-in. The maximum-likelihood model is shown in Figure \ref{fig:wfc3lcFitBroadRaw} and the systematics-corrected light curve with corresponding transit model is shown in Figure \ref{fig:wfc3lcFitBroadCorr}. The root-mean-square of residuals for the maximum-likelihood model is 116\,ppm, equivalent to $2.7 \times$ the photon noise. Inferred model parameter distributions are reported in Table \ref{table:wfc3lcFitBroadRaw}.

\subsection{Spectroscopic transit light curves} \label{sec:lcSpec}

Following the broadband light curve fitting, spectroscopic light curves were produced for 28 equal-width channels spanning the 1.126-1.644$\um$ wavelength range using the methodology described in \cite{2021AJ....161...18M}. In brief, a lateral shift in wavelength and a vertical stretch in flux were applied to the individual 1D spectra to minimise the residuals relative to a template spectrum. The latter was chosen to be the 1D spectrum extracted from the final exposure, as it was used to determine the wavelength solution (Section \ref{sec:dataredWFC3}). The resulting residuals were then binned into the 28 wavelength channels. To produce the final light curves, transit signals were added to these time series, with the same properties as those derived from the broadband light curve fit (Section \ref{sec:lcBroad}), but with limb darkening coefficients fixed to values appropriate for the wavelength range of each spectroscopic channel, determined in the same manner described above for the broadband light curve. Using this process, common-mode systematics were effectively removed from the final spectroscopic light curves, as well as systematics associated with pointing drifts along the dispersion axis of the detector.

Fitting of the common-mode-corrected spectroscopic light curves proceeded in a similar manner to the broadband light curve fitting. The only differences were that HST orbital phase $\varphi$ was the only input variable provided for the squared-exponential covariance kernel, and for the mean function the transit signal was multiplied by a $t$-dependent linear trend to capture the remaining systematics not removed by the common-mode correction. As for the broadband light curve fit, white noise rescaling factors $\beta$ were included as free parameters for each spectroscopic light curve, with uniform priors. The only transit parameters allowed to vary in the spectroscopic light curve fits were $\RpRs$ and the quadratic limb darkening coefficients ($u_1$, $u_2$)  for each of the separate spectroscopic light curves. Uniform priors were adopted for the $\RpRs$ values and tight Gaussian priors obtained using PyLDTK were adopted for the $u_1$ and $u_2$ values, as described in Section \ref{sec:lcBroad}. Alternative treatments of the stellar limb darkening were found to have a negligible effect on the final results and are described in Appendix \ref{app:limbDarkening}. Values for $T_c$, $\aRs$, and $b$ were held fixed to the maximum-likelihood values determined from the broadband light curve fit. 

\begin{table}
\begin{minipage}{\columnwidth}
  \centering
\caption{Spectroscopic light curve fit results.  \label{table:wfc3lcFitSpec}}

\begin{tabular}{ccc}

\hline Wavelength & $\RpRs$ & $\left( \RpRs \right)^2$ \\
 ($\mu$m) &  &  (ppm) \\ \hline
1.126-1.144 & $0.05299_{-0.00094}^{+0.00097}$ & $2808_{-99}^{+104}$ \\ 
1.144-1.163 & $0.05305_{-0.00075}^{+0.00073}$ & $2815_{-79}^{+78}$ \\ 
1.163-1.181 & $0.05239_{-0.00071}^{+0.00071}$ & $2745_{-74}^{+75}$ \\ 
1.181-1.200 & $0.05215_{-0.00077}^{+0.00077}$ & $2720_{-80}^{+81}$ \\ 
1.200-1.218 & $0.05270_{-0.00061}^{+0.00059}$ & $2777_{-64}^{+62}$ \\ 
1.218-1.237 & $0.05169_{-0.00077}^{+0.00069}$ & $2671_{-79}^{+72}$ \\ 
1.237-1.255 & $0.05226_{-0.00066}^{+0.00069}$ & $2731_{-69}^{+73}$ \\ 
1.255-1.274 & $0.05187_{-0.00073}^{+0.00066}$ & $2691_{-75}^{+69}$ \\ 
1.274-1.292 & $0.05215_{-0.00050}^{+0.00055}$ & $2720_{-51}^{+58}$ \\ 
1.292-1.311 & $0.05341_{-0.00059}^{+0.00061}$ & $2853_{-63}^{+66}$ \\ 
1.311-1.329 & $0.05187_{-0.00061}^{+0.00066}$ & $2691_{-63}^{+69}$ \\ 
1.329-1.348 & $0.05297_{-0.00066}^{+0.00068}$ & $2806_{-70}^{+72}$ \\ 
1.348-1.366 & $0.05456_{-0.00071}^{+0.00065}$ & $2976_{-77}^{+71}$ \\ 
1.366-1.385 & $0.05433_{-0.00063}^{+0.00065}$ & $2952_{-68}^{+71}$ \\ 
1.385-1.403 & $0.05360_{-0.00065}^{+0.00060}$ & $2873_{-69}^{+65}$ \\ 
1.403-1.422 & $0.05273_{-0.00061}^{+0.00059}$ & $2780_{-64}^{+62}$ \\ 
1.422-1.440 & $0.05325_{-0.00071}^{+0.00068}$ & $2836_{-75}^{+73}$ \\ 
1.440-1.459 & $0.05450_{-0.00053}^{+0.00051}$ & $2971_{-58}^{+55}$ \\ 
1.459-1.477 & $0.05236_{-0.00078}^{+0.00078}$ & $2741_{-81}^{+82}$ \\ 
1.477-1.496 & $0.05359_{-0.00062}^{+0.00062}$ & $2872_{-66}^{+67}$ \\ 
1.496-1.514 & $0.05246_{-0.00080}^{+0.00077}$ & $2752_{-83}^{+81}$ \\ 
1.514-1.533 & $0.05286_{-0.00069}^{+0.00078}$ & $2794_{-73}^{+83}$ \\ 
1.533-1.551 & $0.05280_{-0.00084}^{+0.00078}$ & $2788_{-88}^{+83}$ \\ 
1.551-1.570 & $0.05248_{-0.00059}^{+0.00057}$ & $2754_{-61}^{+60}$ \\ 
1.570-1.588 & $0.05213_{-0.00072}^{+0.00068}$ & $2718_{-74}^{+71}$ \\ 
1.588-1.607 & $0.05318_{-0.00065}^{+0.00066}$ & $2828_{-69}^{+71}$ \\ 
1.607-1.625 & $0.05253_{-0.00066}^{+0.00068}$ & $2759_{-69}^{+72}$ \\ 
1.625-1.644 & $0.05239_{-0.00052}^{+0.00052}$ & $2745_{-54}^{+55}$ \\ 
\end{tabular}
\end{minipage}
\end{table}

\begin{figure*}
\centering  
\includegraphics[width=\linewidth]{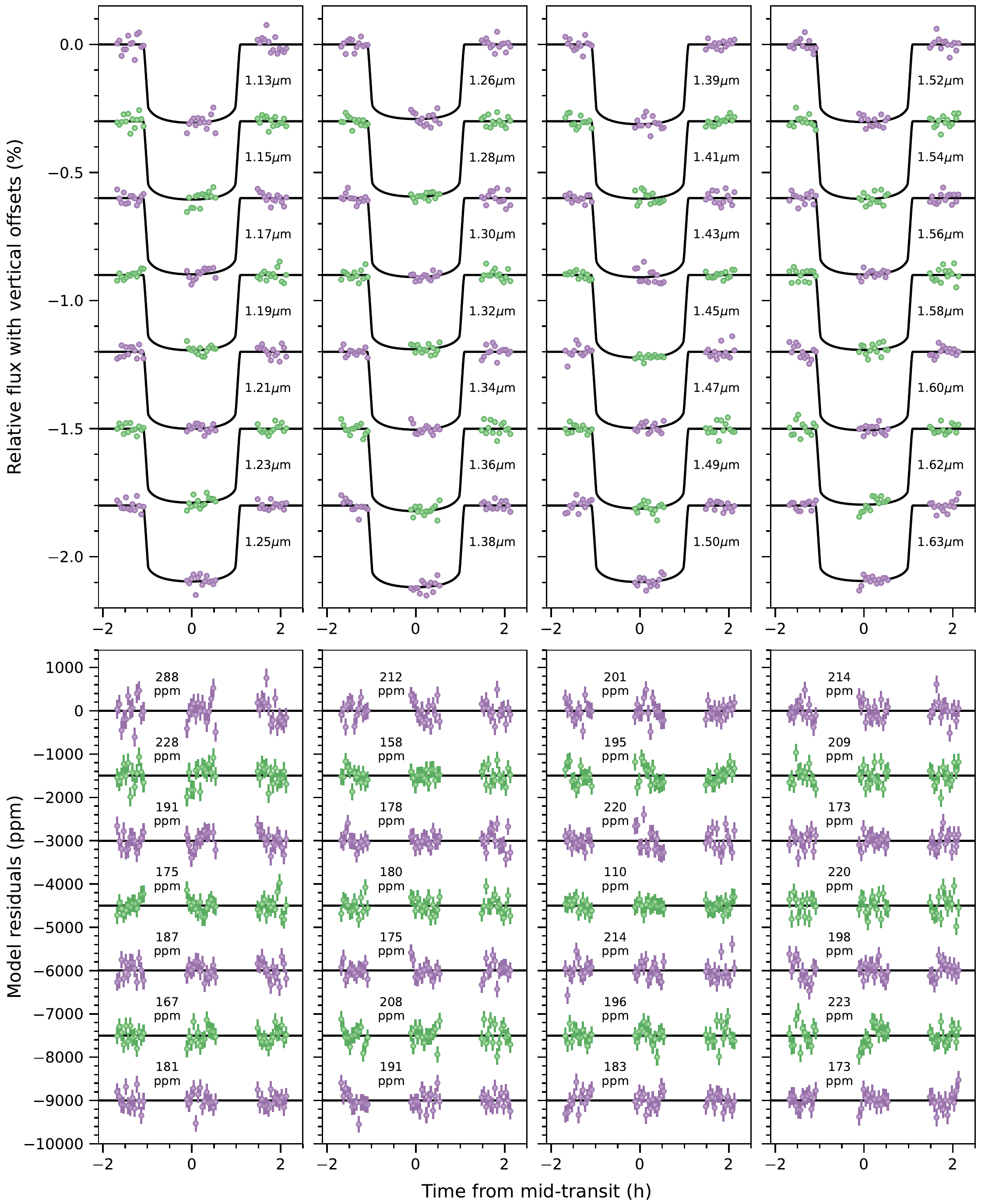}
\caption{(Top panels) Spectroscopic light curves after dividing through by the systematics component of the maximum likelihood light curve models. Data are plotted using alternating colours for visual clarity. (Bottom panels) Maximum likelihood model residuals for each spectroscopic light curve, with corresponding RMS values printed. Note that the plotted error bars are the $1\sigma$ photon noise uncertainties, but the white noise values were treated as free parameters in the model fits as described in the text.}
\label{fig:wfc3lcFitSpec}
\end{figure*}

\begin{figure}
\centering  
\includegraphics[width=\linewidth]{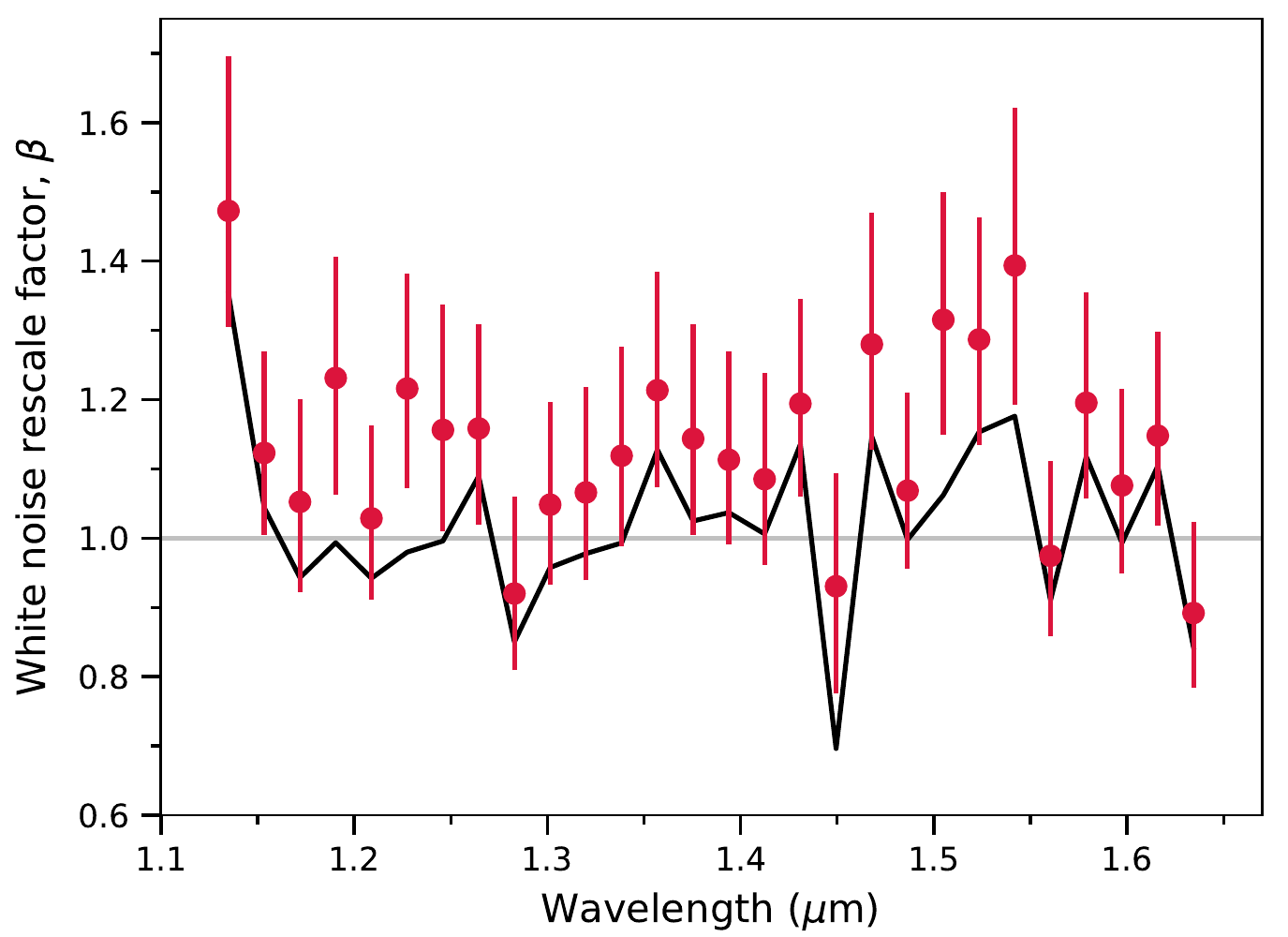}
\caption{Inferred white noise rescaling factors for each spectroscopic light curve. Red points show median values with errorbars giving the 68\% credible intervals. The black line shows the maximum likelihood values.}
\label{fig:wfc3wnoiseSpec}
\end{figure}

Inferred values for $\RpRs$ and the corresponding transit depths $(\RpRs)^2$ are reported in Table \ref{table:wfc3lcFitSpec}. A median precision of 71\,ppm is achieved for the transit depth measurements across all wavelength channels. Systematics-corrected light curves are shown in Figure \ref{fig:wfc3lcFitSpec} and the inferred white noise rescaling factors are plotted in Figure \ref{fig:wfc3wnoiseSpec}. Posterior distributions for the limb darkening coefficients are listed in Table \ref{table:wfc3lcSpecLD} and are effectively identical to the adopted PyLDTK priors (Appendix \ref{app:limbDarkening}).

\section{STIS analysis: stellar Ly$\alpha$} \label{sec:dataredSTIS}

\begin{figure}
\centering  
\includegraphics[width=0.9\columnwidth]{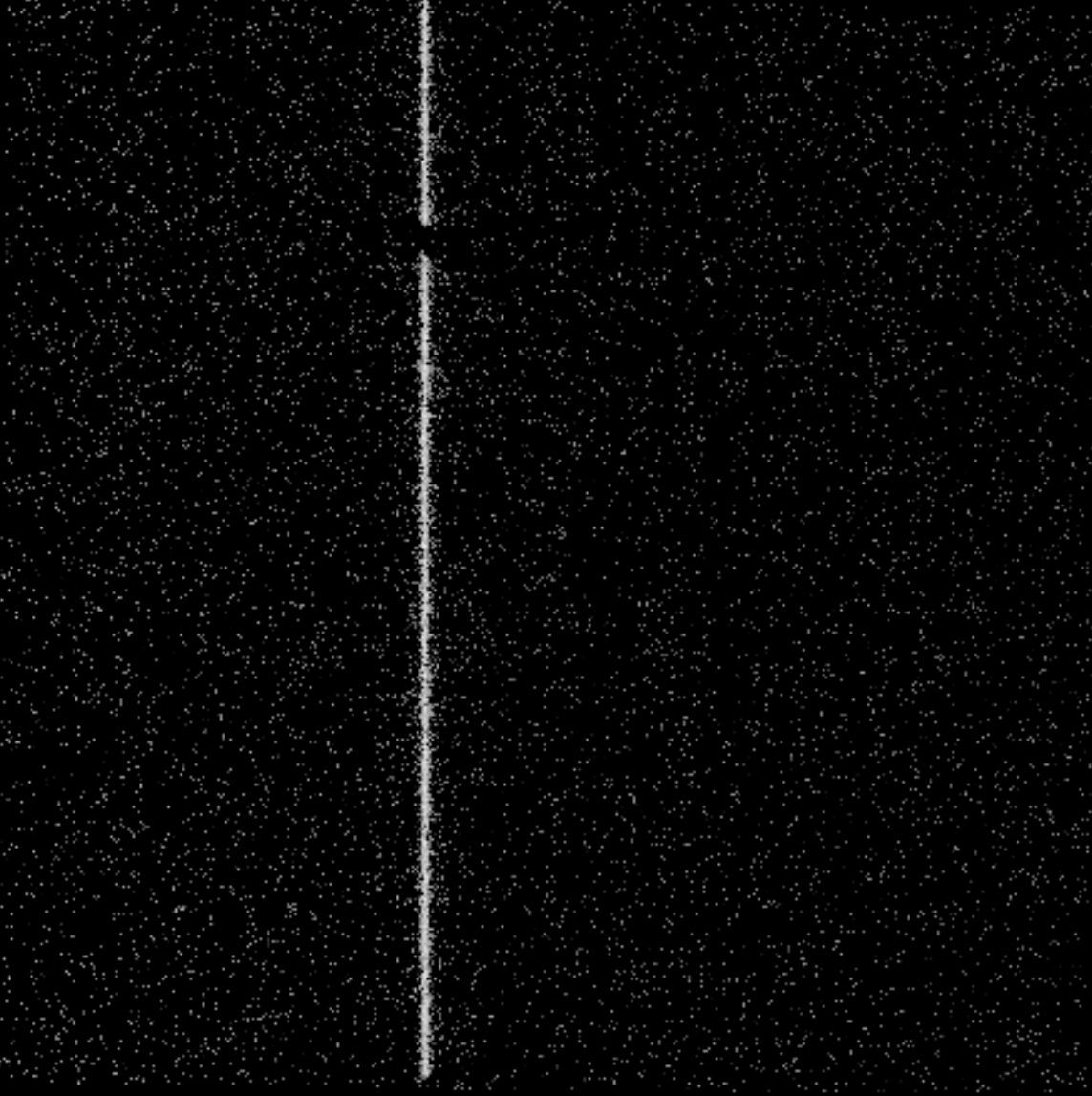}
\includegraphics[width=0.9\columnwidth]{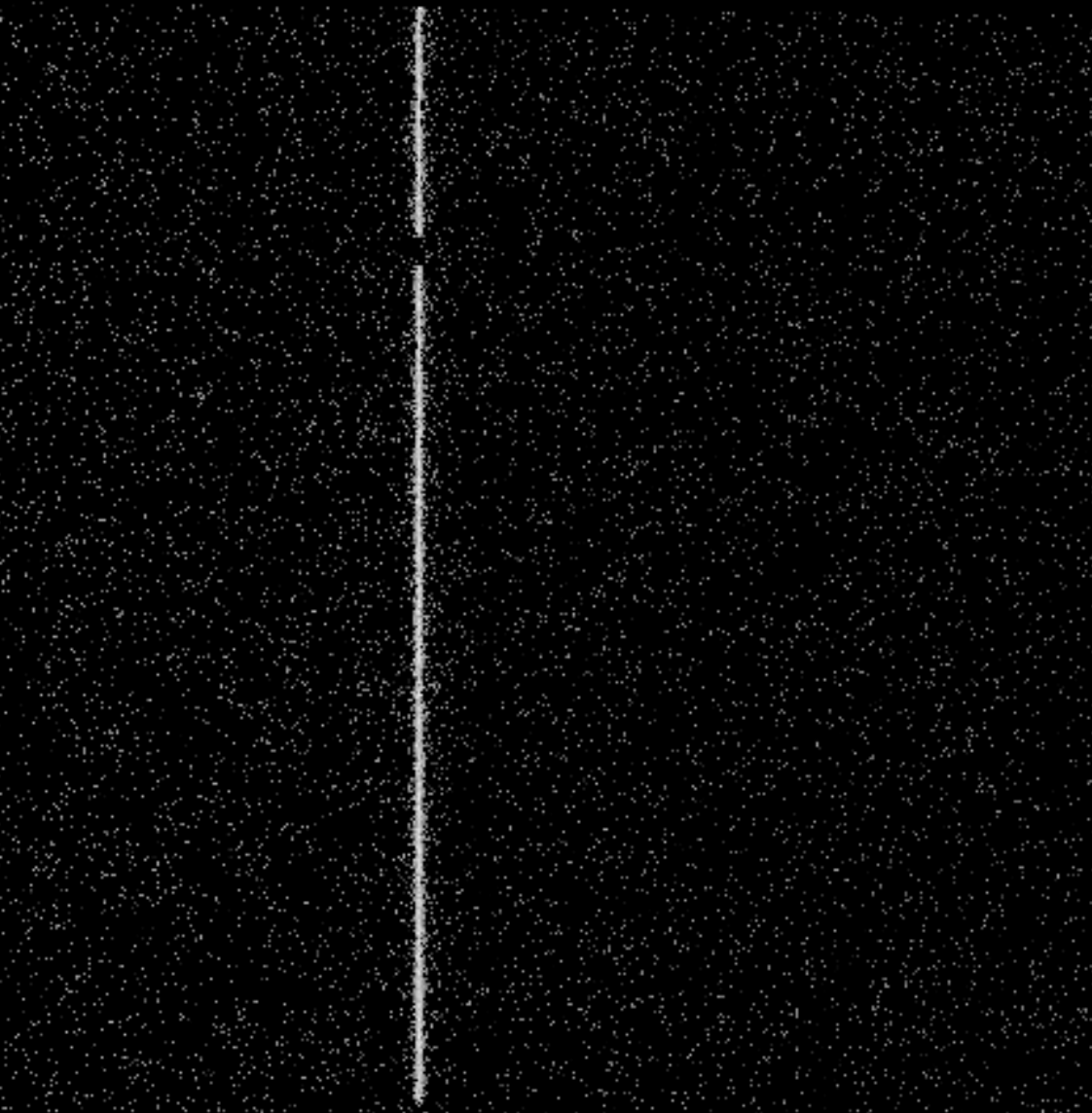}
\caption{Calibrated HST STIS Ly$\alpha$ observations of TOI-270. \textit{(Top)} Calibrated spectrum taken on 2020 January 31. Wavelength dispersion direction is along the horizontal axis. The bright stripe down the middle is the Earth's Ly$\alpha$ geocoronal emission. Ly$\alpha$ emission from TOI-270 would be visible alongside this line at the Doppler shift corresponding to the velocity difference between HST and the target. We find no significant evidence of Ly$\alpha$ emission in this exposure. \textit{(Bottom)} Identical as above, but for a second exposure taken on 2020 January 29. }
\label{fig:STIS_UV}
\end{figure}

The STIS G140M data were reduced using the standard STIS pipeline (\texttt{CALSTIS} version 3.4.2). Figure \ref{fig:STIS_UV} shows the reduced and calibrated two-dimensional spectra. The geocoronal Ly$\alpha$ emission is clearly visible as a strong vertical stripe (the horizontal axis is the dispersion direction). Emission from TOI-270 would lie adjacent to the geocoronal emission due to the Doppler shift between HST and TOI-270. However, no evidence is visible in either exposure for emission from TOI-270.

\begin{figure}
\centering  
\includegraphics[width=0.9\columnwidth]{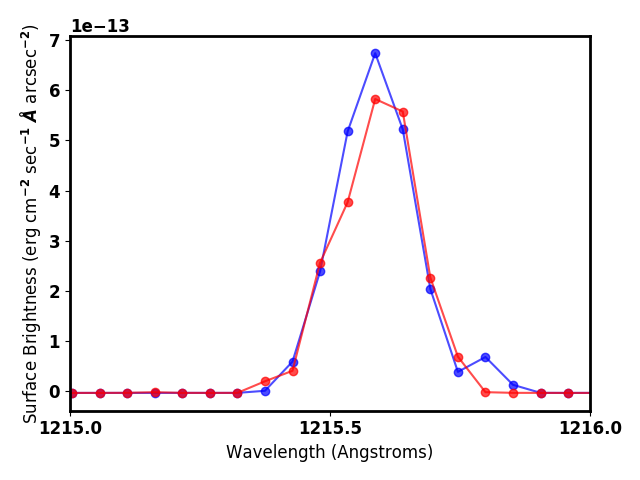}
\caption{Ly$\alpha$ emission of TOI-270 for the exposure on 2020-01-29 (red) and 2020-01-31 (blue). The large emission line is due to the Earth's geocoronal emission. The slight bump redwards of this line on the 2020-01-31 data set is suggestive of brief Ly$\alpha$ emission of TOI-270. However, we do not recover this emission on the other observation. Additionally, we see comparable noise bursts in other areas of our 2-D spectrum (Figure \ref{fig:STIS_UV}), and therefore we suggest that we have not detected Ly$\alpha$ emission from TOI-270.}
\label{fig:Lyalpha}
\end{figure}

To assess the sensitivity of the STIS data to Ly$\alpha$ emission from TOI-270, a spectrum was extracted (including the geocoronal emission) at the target location for each exposure. These data are shown in Figure \ref{fig:Lyalpha}. We note that during the second observation (2020 January 31), there is an apparent flux excess redwards of the geocoronal emission line. However, there is no evidence for this excess in the first observation (2020 January 29). Furthermore, the derived error bar at this wavelength is $ 0.6 \times 10^{-13}$ erg cm$^{-2}$ sec$^{-1}$ \AA$^{-1}$ arcsec$^{-2}$. Given that the amplitude of the excess is only about  $10^{-13}$ erg cm$^{-2}$ sec$^{-1}$ \AA$^{-1}$ arcsec$^{-2}$, this translates to a $<2\sigma$ detection significance. Therefore we do not believe this feature in the data to be due to significant emission from TOI-270. 

We derive a conservative upper-limit for the stellar Ly$\alpha$ emission from the data errorbars close to where confusion with the Earth's emission becomes significant, and estimate that we could have detected line emission peaking approximately twice as high as these error bars, and spread over the resolving power of the instrument. Using this approach, we place a $95\%$ credible upper limit of $8.2 \times 10^{-14}$\,erg\,s$^{-1}$\,cm$^2$\,\AA$^{-1}$\,arcsec$^{-2}$ on the Ly$\alpha$ emission of TOI-270. 

\section{Atmospheric retrieval analysis} \label{sec:retrieval}

We use the transmission spectrum of TOI-270\,d to retrieve the atmospheric properties at the day-night terminator region of the planet using the AURA atmospheric retrieval framework for transmission spectra \citep{pinhas2018}. The model assumes a plane-parallel atmosphere in hydrostatic equilibrium. The temperature structure, chemical composition, and the properties of clouds/hazes are free parameters in the model. We consider opacity contributions due to prominent chemical species expected in temperate hydrogen-rich atmospheres in chemical equilibrium and disequilibrium, which include H$_2$O, CH$_4$, NH$_3$, CO, CO$_2$, and HCN \citep[e.g.][]{Madhu2011,Moses2013,Yu2021,Hu2021,Tsai2021}. The opacities of the molecules were obtained using corresponding line lists from the HITRAN and Exomol databases: H$_2$O \citep{rothman2010}, CH$_4$ \citep{yurchenko2014}, NH$_3$ \citep{yurchenko2011},  CO \citep{rothman2010}, CO$_2$ \citep{rothman2010}, and HCN \citep{barber2014}, and collision-induced absorption (CIA) due to H$_2$-H$_2$ and H$_2$-He \citep{richard2012}. The absorption cross sections are computed from the line lists following \citet{gandhi2017}. Besides molecular absorption, we also consider Rayleigh scattering due to H$_2$ and parametric contributions from clouds/hazes and stellar heterogeneity as described in \cite{pinhas2018}. The Bayesian inference and parameter estimation in the retrievals are conducted using the Nested Sampling algorithm  \citep{Feroz2009} implemented with the PyMultiNest package \citep{Buchner2014}. 

\begin{figure*}
\centering  
\includegraphics[width=\linewidth]{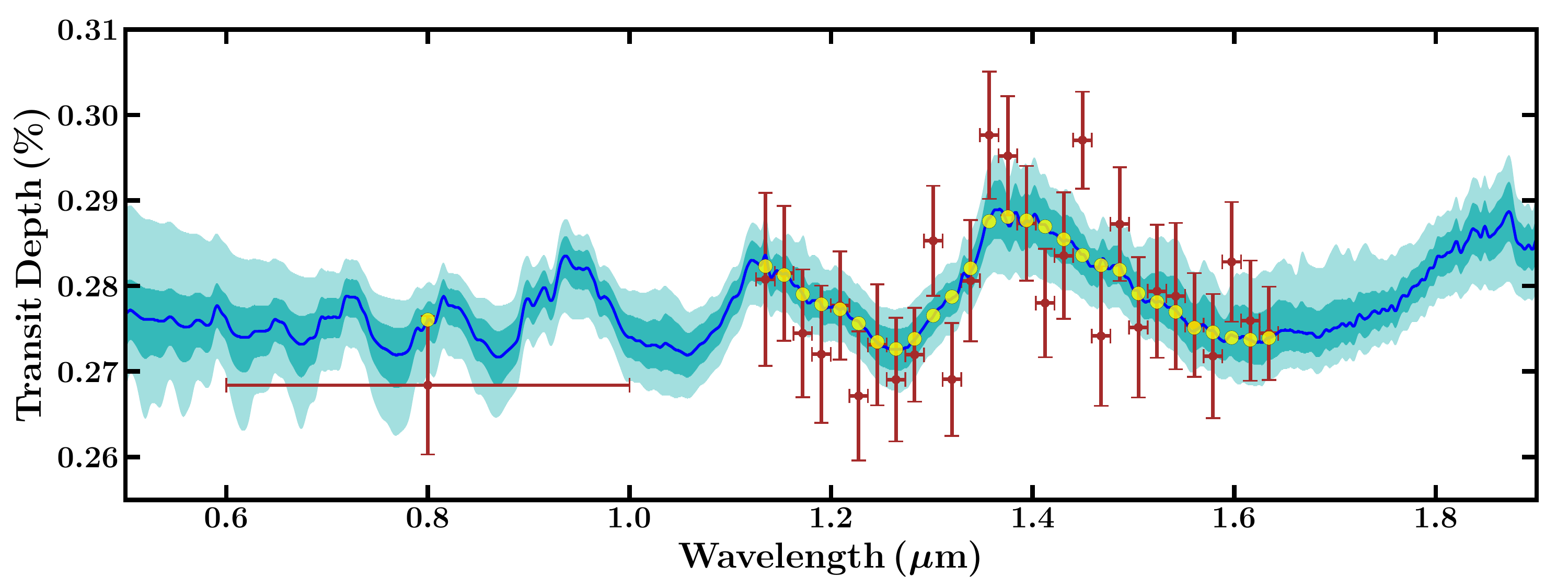}
\includegraphics[width=\linewidth]{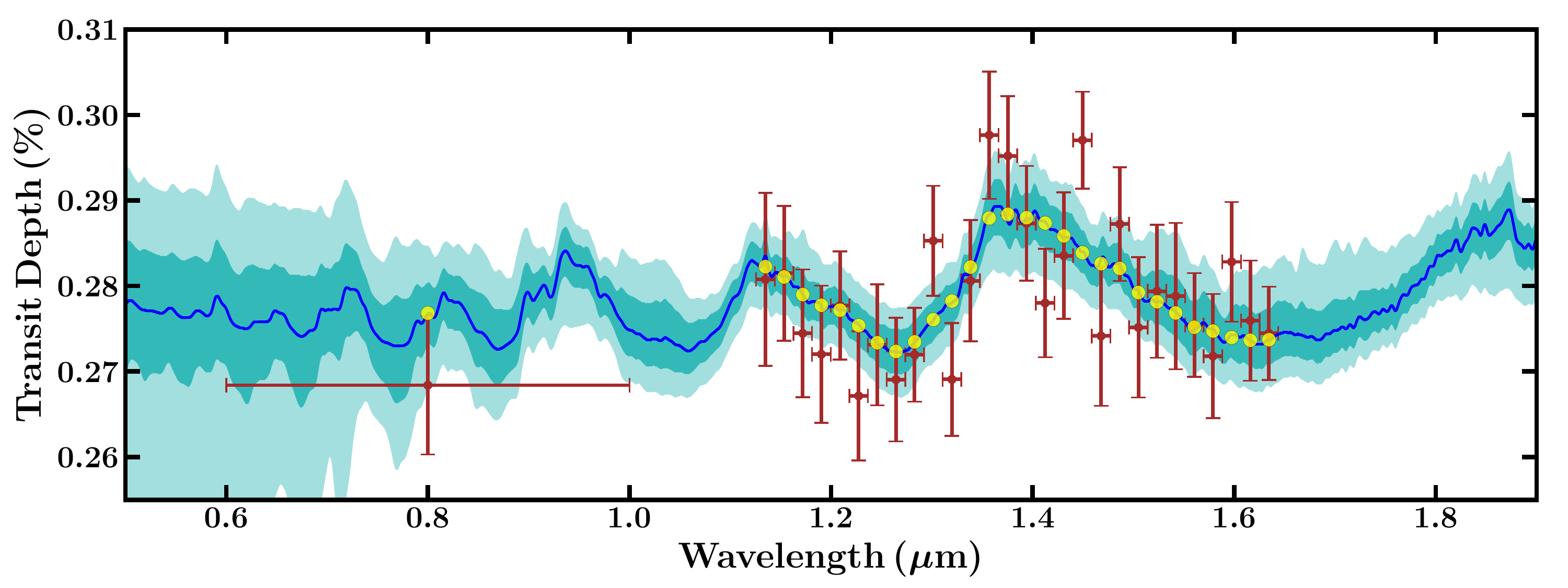}
\caption{Observations and retrieval of the transmission spectrum of TOI-270\,d. Top: Retrieval with baseline planetary atmosphere model and no stellar heterogeneity. Bottom: Retrieval including stellar heterogeneity. Both retrievals included the TESS (photometric point at 0.8 $\mu$m) and WFC3 (spectrum between 1.1-1.7 $\mu$m) data, shown in red. The blue curve shows the median-fit model spectrum and the light blue shaded regions show the $1\sigma$ and $2\sigma$ credible ranges. The binned model points for the median-fit spectrum are shown as yellow circles.}
\label{fig:spectrum}
\end{figure*}

\begin{figure*}
\centering  
\includegraphics[width=\linewidth]{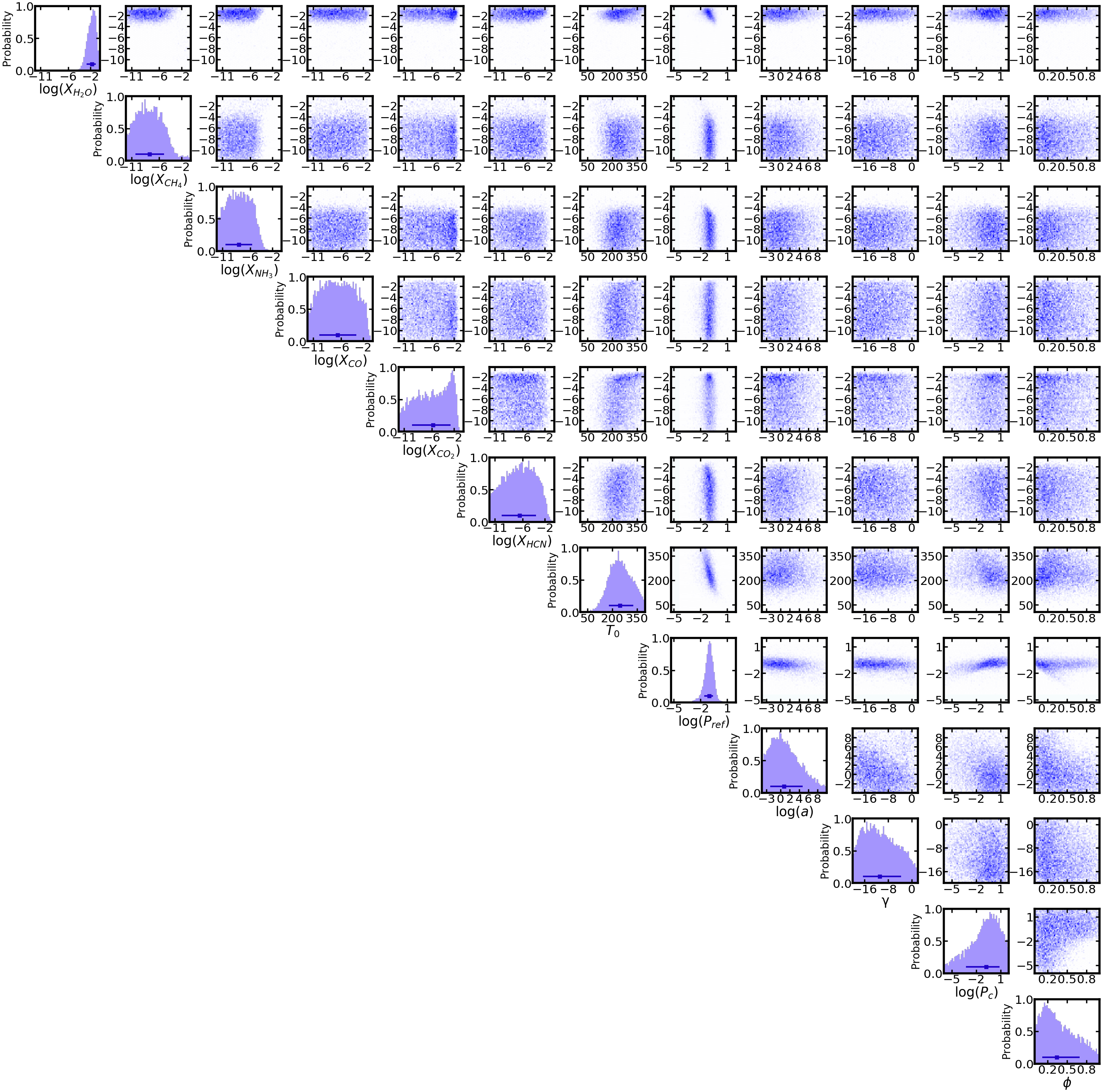}
\caption{Posterior probability distributions for the atmospheric model parameters of TOI-270\,d obtained from the retrieval without stellar heterogeneity and including both the TESS and WFC3 data. The inset at the bottom left shows the retrieved values, for the volume mixing ratios of the molecules ($X_{\rm i}$), isothermal temperature ($T_0$), reference pressure in bar ($P_{\rm ref}$), haze amplitude and slope ($a$ and $\gamma$), cloud-top pressure in bar ($P_c$) and cloud/haze covering fraction ($\bar{\phi}$). The retrievals show moderate evidence for H$_2$O at 3.0 $\sigma$ significance.  No other molecules are individually detected above 2 $\sigma$, and the model does not show any evidence for clouds/hazes in the observed slant photosphere. The retrieved parameters are reported in Table~\ref{table:priors}.}
\label{fig:posteriors}
\end{figure*}

\subsection{Baseline retrieval assuming uniform stellar brightness} \label{sec:retrievalBaseline}

We follow a similar approach to the retrieval conducted using AURA for the temperate sub-Neptune K2-18\,b \citep[][]{Madhu2020}. We consider an isothermal pressure-temperature ($P$-$T$) profile with the temperature as a free parameter. Assuming a non-isothermal profile does not significantly influence the results and is not warranted given the data quality \citep[also see e.g.][]{Constantinou2022}. Our baseline model, therefore, includes twelve free parameters: volume mixing ratios of the six molecules noted above ($X_{\rm i}$ for $i \ =$~{H$_2$O, CH$_4$, NH$_3$, CO, CO$_2$, HCN}); the isothermal temperature ($T_0$); the reference pressure ($P_{\rm ref}$) corresponding to the planet radius of 2.19 R$_\oplus$ derived from the broadband light curve (Table~\ref{table:wfc3lcFitBroadRaw}); and four cloud/haze parameters, including the haze amplitude and slope ($a$ and $\gamma$), cloud-top pressure ($P_c$) and cloud/haze covering fraction ($\phi$). The model fit to the observed spectrum is shown in Figure \ref{fig:spectrum} and the posterior probability distributions for the model parameters are shown in Figure \ref{fig:posteriors}.

For this baseline model, we find a preference for absorption by a combination of the molecular species listed above in a hydrogen-rich atmosphere at 4$\sigma$ significance. The significance is derived by comparing the Bayesian evidence  \citep[e.g.,][]{Benneke2013,Welbanks2021} for the baseline model relative to that obtained for a featureless spectrum, i.e., a flat line. When considered individually, the evidence in favour of a given molecule being present in the atmosphere is lower, with the strongest evidence being obtained for H$_2$O at $3\sigma$ significance. We do not find significant evidence ($>2\sigma$) for absorption due to any other individual molecules. The retrieved parameters are reported in Table~\ref{table:priors}.

Using the baseline model we retrieve the H$_2$O abundance to be $\log( \XHtwoO ) =  -1.77^{+0.69}_{-0.93}$.  We do not obtain strong constraints on the volume mixing ratios for any of the other molecules, as shown in Figure \ref{fig:posteriors}. The cloud/haze properties are also poorly constrained, with the cloud-top pressure retrieved as $\log(P_{\rm c}/{\rm bar}) = -0.77^{+1.63}_{-2.47}$ and a cloud/haze covering fraction of $0.38^{+0.33}_{-0.23}$, i.e. consistent with a cloud-free slant photosphere at $2\sigma$. The isothermal temperature is constrained to $247^{+80}_{-68}$ K.

To investigate the extent to which the results described above are driven by the relative transit depth levels of the TESS and WFC3 passbands, a second retrieval was performed using only the WFC3 data, with the TESS data point excluded. Similar to the above case, we still find evidence for H$_2$O in the planetary atmosphere at 3$\sigma$ significance, with no significant constraints on any other absorber or clouds/hazes. An H$_2$O abundance of $\log(\XHtwoO) =  -1.92^{+0.74}_{-0.97}$ is obtained, which is fully consistent with the constraint obtained from the original analysis that included the TESS data point.

\subsection{Retrievals allowing for stellar heterogeneity} \label{sec:retrievalStellarHeterogeneity}

We also investigate the possibility of a heterogeneous stellar brightness profile influencing the observed transmission spectrum, as has been suggested for other temperate sub-Neptune atmospheres \citep[e.g.][]{rackham2018,2021AJ....162..300B}. We include the effect of stellar heterogeneity following \citet{pinhas2018}, by adding three parameters to the baseline retrieval described in the previous section: the average stellar photospheric temperature ($T_{\rm phot}$); the average temperature of the heterogeneity ($T_{\rm het}$); and the covering fraction of the heterogeneity ($\delta$). When considering the WFC3 and TESS data together and allowing for stellar heterogeneity, we find that the retrieved planetary atmosphere properties do not change significantly, and the quality of the fit to the data is similar to that obtained for the retrievals that assumed a homogeneous stellar brightness profile. With stellar heterogeneity included, we obtain evidence for H$_2$O absorption in the planetary atmosphere at 3$\sigma$ significance, with a derived abundance of $\log(\XHtwoO) =  -1.91^{+0.74}_{-0.98}$, fully consistent with the abundances obtained for the retrievals without stellar heterogeneity. Again, we find no significance evidence for any other chemical species in the planetary atmosphere. For the star spots, we infer a covering fraction of $\delta =0.06^{+0.05}_{-0.03}$ and temperature of $T_{\rm het}=2446^{+1120}_{-988}$ K. The resulting model fit is shown in Figure \ref{fig:spectrum}.

When considering the combined TESS and WFC3 dataset, the baseline model without stellar heterogeneity is favoured over the model with stellar heterogeneity at $2\sigma$ significance. In both cases, the fit to the data is primarily driven by the H$_2$O absorption in the planetary atmosphere. The model with stellar heterogeneity is marginally disfavored due to the larger number of free parameters that do not contribute significantly to the fit. In particular, a substantial contamination by stellar heterogeneities would produce a larger offset between the TESS and WFC3 transit depth levels than that which is observed \citep{pinhas2018}.

\begin{figure*}
\centering  
\includegraphics[width=\linewidth]{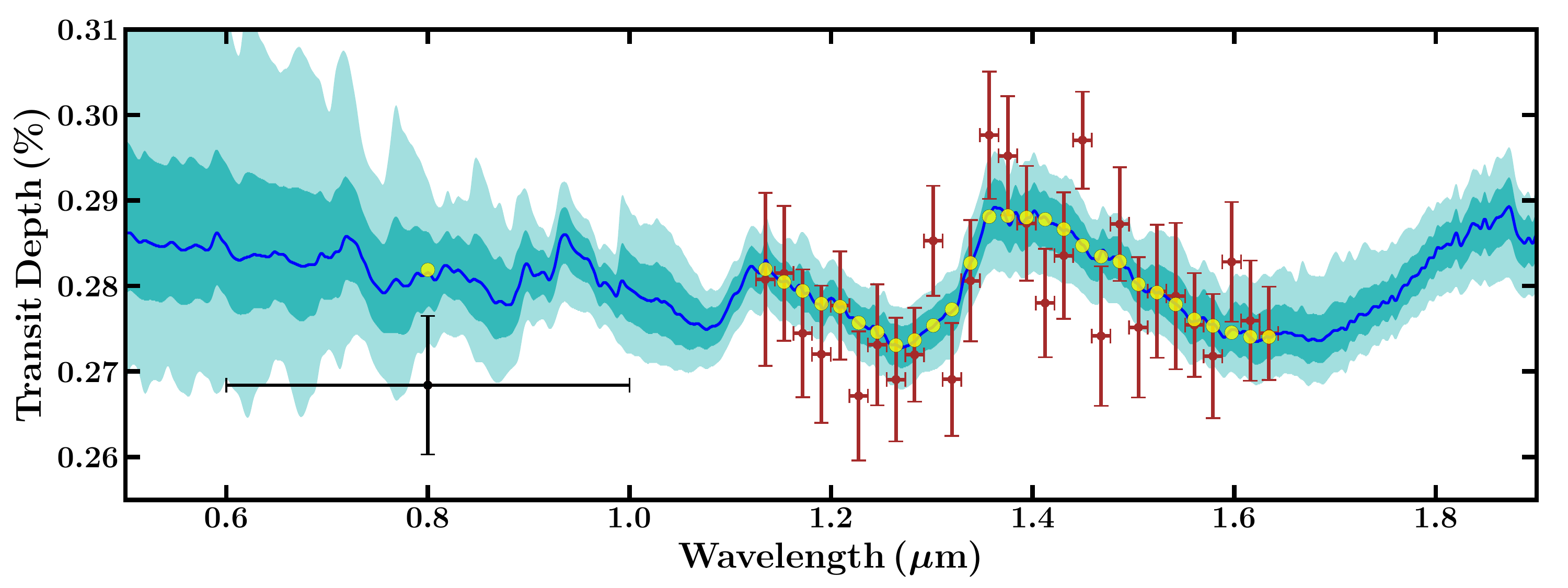}
\caption{Retrieved transmission spectrum considering WFC3 data alone and a model including stellar heterogeneity. The blue curve shows the median-fit model spectrum and the light blue shaded regions show the $1\sigma$ and $2\sigma$ credible ranges. The binned model points for the median-fit spectrum are shown as yellow circles. The TESS data point at 0.8 $\mu$m is not included in the retrieval but is shown here for comparison.}
\label{fig:spectrum_wfc3only}
\end{figure*}

To determine the information content of the WFC3 data alone, we performed an additional retrieval including stellar heterogeneity but with the TESS data point excluded. The resulting model fit is shown in Figure \ref{fig:spectrum_wfc3only}. In this case, we find that the detection significance of H$_2$O is reduced to 2.8$\sigma$ with an abundance constraint of $\log( \XHtwoO ) = -2.09^{+0.89}_{-1.21}$, which remains consistent with the abundance constraints obtained from the retrievals that included the TESS data point. Comparing the Bayesian evidences for WFC3-only retrievals with and without stellar heterogeneity, we find that there is still no strong preference for contamination by stellar heterogeneity; the evidence remains marginally higher for the baseline model without stellar heterogeneity. As an additional check, we performed a retrieval on the WFC3 data alone assuming a model with a featureless planetary spectrum and a heterogeneous stellar brightness profile. We find that this scenario is also disfavored at over $3\sigma$ significance relative to the baseline retrieval allowing for a planetary atmosphere and not including stellar heterogeneity. Nevertheless, the marginal detection significance (2.8$\sigma$) obtained for H$_2$O when allowing for stellar heterogeneity and considering the WFC3 data alone, and our current inability to definitively rule out contamination by stellar heterogeneity, implies that more observations are required to robustly establish the presence of H$_2$O in the planetary atmosphere.

\section{Discussion} \label{sec:discussion}

Although we cannot yet definitively discount the possibility that the measured transmission spectrum is contaminated by stellar heterogeneity, various indicators point to the TOI-270 host star being a particularly quiet M dwarf, likely with an age of at least a billion years. For example, \cite{Guenther2019} report two measurements that are suggestive of a very low stellar activity level. First, they find an absence of fast rotational modulation, transit spot crossings, and flares in existing TESS photometry, which spans nearly three years \citep[non-continuous; see also][]{Kaye2022}. Second, the same authors detected a shallow H$\alpha$ absorption feature and reported an absence of Ca H/K emission lines in their reconnaissance spectra. Both findings are common signs for less active M dwarfs, while active stars would show H$\alpha$ and Ca H/K in emission \citep[e.g.][]{Cram1985, Stauffer1986, Robertson2013}. A subsequent study by \citet{VanEylen2021} reports a tentative stellar rotation period of $\sim 57$ days derived from a periodogram analysis of TESS photometry, matching the expectations for an old, quiet M dwarf \citep{Newton2016}. Van Eylen et al.\ additionally analyse HARPS and ESPRESSO radial velocity measurements, which independently confirm very low stellar activity. For example, the ESPRESSO radial velocity jitter is constrained to $0.16 \pm 0.23$\,m\,s$^{-1}$, consistent with zero. Additionally, the lack of Ly$\alpha$ emission detected in the present study (Section \ref{sec:dataredSTIS}) is consistent with a low activity level for the host star. The low stellar activity level suggested by these various auxiliary indicators are in line with the preference we obtain for the measured transmission spectrum being minimally contaminated by stellar heterogeneity (Section \ref{sec:retrieval}).

Pending further observations, if we assume for now that the measured WFC3 signal is caused by the atmosphere of TOI-270\,d rather than unocculted star spots, then the retrieved chemical abundances provide initial constraints on the bulk planetary composition and surface conditions. TOI-270\,d has been predicted to be a candidate Hycean world \citep{2021ApJ...918....1M}, i.e. an ocean world with a hydrogen-rich atmosphere capable of hosting a habitable ocean. The mass ($4.20 \pm 0.16 \ME$), broadband radius ($2.17 \pm 0.05 \, \RE$), and equilibrium temperature (350 K) of the planet together allow for a range of internal structures, including (a) a gas dwarf with a rocky core overlaid by a thick hydrogen-rich atmosphere, (b) a water world with a steamy atmosphere ($\sim$80-100\% H$_2$O volume mixing ratio), and (c) a Hycean world composed of a water-rich interior and a hydrogen-rich atmosphere. In this study, we obtain a 99\% credible upper limit of 30\% for the H$_2$O volume mixing ratio, thereby precluding a water world with a steamy atmosphere composed of $>80\%$ H$_2$O. However, more stringent constraints on the atmospheric molecular abundances will be needed to robustly distinguish between the gas dwarf versus Hycean world scenarios.  

The retrieved H$_2$O abundance provides a tentative constraint on the atmospheric metallicity of the planet. The H$_2$O abundance we retrieve for TOI-270~d, $\log(\XHtwoO) =  -1.77^{+0.69}_{-0.93}$, is similar to that reported previously for the sub-Neptune K2-18~b \citep{Madhu2020}, $\log(\XHtwoO) =  -2.11^{+1.06}_{-1.19}$. Assuming all the oxygen in the atmosphere is in H$_2$O, the retrieved H$_2$O abundance of TOI-270~d corresponds to an O/H ratio spanning 2-99$\times$ solar, with a median value of $20\times$ solar. This estimate is consistent with the mass-metallicity trend observed for  close-in exoplanets using oxygen as a metallicity indicator \citep{Welbanks2019} as well as with theoretical predictions of atmospheric metal enrichment in low-mass exoplanets, albeit at the lower-end of the theoretically predicted range \citep[e.g.][]{Fortney2013}.

The derived atmospheric abundances also potentially provide insights into chemical processes in the atmosphere, which in turn could constrain the surface conditions. In chemical equilibrium, given the low temperature ($\sim$300-350 K) of the atmosphere, CH$_4$ and NH$_3$ are expected to be the prominent carbon and nitrogen bearing species, respectively \citep{Madhu2011,Moses2013}, similar to that seen in the solar system giant planets. Assuming solar elemental ratios of C/O and N/O, the abundances of CH$_4$ and NH$_3$ are expected to be $\sim$0.5$\times$ and $\sim$0.2$\times$ the H$_2$O abundance, respectively. However, while H$_2$O is retrieved in abundance, as discussed above, we do not retrieve comparable constraints for CH$_4$ and NH$_3$. The lack of significant evidence for CH$_4$ and NH$_3$ may indicate chemical disequilibrium in the atmosphere, as also noted for K2-18\,b \citep{Madhu2020}. 

Such disequilibrium may result if the base of the outer atmospheric envelope occurs at a pressure of less than $\sim$ 100 bar and interfaces with an interior layer of different composition \citep{Yu2021,Hu2021,Tsai2021}. For example, a water ocean below a shallow hydrogen-dominated envelope can deplete CH$_4$ and NH$_3$ in the observable atmosphere. The chemical composition inferred for TOI-270\,d may, therefore, be consistent with the presence of a water ocean below the outer hydrogen-rich envelope, similar to the constraints inferred for the habitable-zone sub-Neptune K2-18\,b \citep{Madhu2020}. However, more precise abundance constraints from future observations will be required to further elucidate the interior and surface conditions of TOI-270\,d.

\section{Conclusion} \label{sec:conclusion}

We have presented an atmospheric transmission spectrum for the temperate sub-Neptune TOI-270\,d measured using the HST WFC3 spectrograph, spanning near-infrared wavelengths 1.126-1.644$\um$. We also used observations made with the HST STIS spectrograph to place a 95\% credible upper limit of $8.2 \times 10^{-14}$\,erg\,s$^{-1}$\,cm$^2$\,\AA$^{-1}$\,arcsec$^{-2}$ on the Ly$\alpha$ emission of the host star. Assuming a homogeneous brightness profile for the stellar disc, the combined TESS and WFC3 transmission spectrum provides strong ($4\sigma$) evidence for infrared absorption by the planetary atmosphere. Although the present data precision make it challenging to uniquely identify the molecules present, the strongest evidence is obtained for H$_2$O at $3\sigma$ significance. If we allow for the possibility of a heterogeneous stellar brightness profile and exclude the TESS point from our analysis, the detection significance of H$_2$O in the planetary atmosphere is reduced to $2.8\sigma$, but the model is disfavoured at the $2\sigma$ level relative to the model assuming a homogeneous stellar brightness profile. Furthermore, even when we allow for stellar heterogeneity and the TESS point is excluded, the detection of a planetary atmosphere is still preferred over a featureless planetary spectrum at more than $3\sigma$ significance.

While the current data broadly favour the detection of a planetary atmosphere and a homogeneous stellar brightness profile, further observations will be required to increase the robustness of the H$_2$O detection and to firmly rule out contamination of the measured signal by stellar heterogeneity. If verified, the inferred atmospheric H$_2$O abundance of $\log( \XHtwoO ) = -1.77^{+0.69}_{-0.93}$ would imply that TOI-270\,d has a hydrogen-rich outer envelope and is a not water world with a steam-dominated atmosphere. The constraints obtained for the atmospheric composition remain compatible with TOI-270\,d being a Hycean planet with a water ocean layer below a hydrogen-rich outer envelope.

\bibliographystyle{apj}
\bibliography{TOI-270d}

\begin{thebibliography}{}
\expandafter\ifx\csname natexlab\endcsname\relax\def\natexlab#1{#1}\fi

\bibitem[{{Almenara} {et~al.}(2022){Almenara}, {Bonfils}, {Otegi}, {Attia},
  {Turbet}, {Astudillo-Defru}, {Collins}, {Polanski}, {Bourrier}, {Hellier},
  {Ziegler}, {Bouchy}, {Briceno}, {Charbonneau}, {Cointepas}, {Collins},
  {Crossfield}, {Delfosse}, {Diaz}, {Dorn}, {Doty}, {Forveille}, {Gaisn{\'e}},
  {Gan}, {Helled}, {Hesse}, {Jenkins}, {Jensen}, {Latham}, {Law}, {Mann},
  {Mao}, {McLean}, {Murgas}, {Myers}, {Seager}, {Shporer}, {Tan}, {Twicken}, \&
  {Winn}}]{2022A&A...665A..91A}
{Almenara}, J.~M., {Bonfils}, X., {Otegi}, J.~F., {et~al.} 2022, \aap, 665, A91

\bibitem[{{Astropy Collaboration} {et~al.}(2013){Astropy Collaboration},
  {Robitaille}, {Tollerud}, {Greenfield}, {Droettboom}, {Bray}, {Aldcroft},
  {Davis}, {Ginsburg}, {Price-Whelan}, {Kerzendorf}, {Conley}, {Crighton},
  {Barbary}, {Muna}, {Ferguson}, {Grollier}, {Parikh}, {Nair}, {Unther},
  {Deil}, {Woillez}, {Conseil}, {Kramer}, {Turner}, {Singer}, {Fox}, {Weaver},
  {Zabalza}, {Edwards}, {Azalee Bostroem}, {Burke}, {Casey}, {Crawford},
  {Dencheva}, {Ely}, {Jenness}, {Labrie}, {Lim}, {Pierfederici}, {Pontzen},
  {Ptak}, {Refsdal}, {Servillat}, \& {Streicher}}]{2013A&A...558A..33A}
{Astropy Collaboration}, {Robitaille}, T.~P., {Tollerud}, E.~J., {et~al.} 2013,
  \aap, 558, A33

\bibitem[{{Astropy Collaboration} {et~al.}(2018){Astropy Collaboration},
  {Price-Whelan}, {Sip{\H{o}}cz}, {G{\"u}nther}, {Lim}, {Crawford}, {Conseil},
  {Shupe}, {Craig}, {Dencheva}, {Ginsburg}, {Vand erPlas}, {Bradley},
  {P{\'e}rez-Su{\'a}rez}, {de Val-Borro}, {Aldcroft}, {Cruz}, {Robitaille},
  {Tollerud}, {Ardelean}, {Babej}, {Bach}, {Bachetti}, {Bakanov}, {Bamford},
  {Barentsen}, {Barmby}, {Baumbach}, {Berry}, {Biscani}, {Boquien}, {Bostroem},
  {Bouma}, {Brammer}, {Bray}, {Breytenbach}, {Buddelmeijer}, {Burke},
  {Calderone}, {Cano Rodr{\'\i}guez}, {Cara}, {Cardoso}, {Cheedella}, {Copin},
  {Corrales}, {Crichton}, {D'Avella}, {Deil}, {Depagne}, {Dietrich}, {Donath},
  {Droettboom}, {Earl}, {Erben}, {Fabbro}, {Ferreira}, {Finethy}, {Fox},
  {Garrison}, {Gibbons}, {Goldstein}, {Gommers}, {Greco}, {Greenfield},
  {Groener}, {Grollier}, {Hagen}, {Hirst}, {Homeier}, {Horton}, {Hosseinzadeh},
  {Hu}, {Hunkeler}, {Ivezi{\'c}}, {Jain}, {Jenness}, {Kanarek}, {Kendrew},
  {Kern}, {Kerzendorf}, {Khvalko}, {King}, {Kirkby}, {Kulkarni}, {Kumar},
  {Lee}, {Lenz}, {Littlefair}, {Ma}, {Macleod}, {Mastropietro}, {McCully},
  {Montagnac}, {Morris}, {Mueller}, {Mumford}, {Muna}, {Murphy}, {Nelson},
  {Nguyen}, {Ninan}, {N{\"o}the}, {Ogaz}, {Oh}, {Parejko}, {Parley}, {Pascual},
  {Patil}, {Patil}, {Plunkett}, {Prochaska}, {Rastogi}, {Reddy Janga},
  {Sabater}, {Sakurikar}, {Seifert}, {Sherbert}, {Sherwood-Taylor}, {Shih},
  {Sick}, {Silbiger}, {Singanamalla}, {Singer}, {Sladen}, {Sooley},
  {Sornarajah}, {Streicher}, {Teuben}, {Thomas}, {Tremblay}, {Turner},
  {Terr{\'o}n}, {van Kerkwijk}, {de la Vega}, {Watkins}, {Weaver}, {Whitmore},
  {Woillez}, {Zabalza}, \& {Astropy Contributors}}]{2018AJ....156..123A}
{Astropy Collaboration}, {Price-Whelan}, A.~M., {Sip{\H{o}}cz}, B.~M., {et~al.}
  2018, \aj, 156, 123

\bibitem[{{Barber} {et~al.}(2014){Barber}, {Strange}, {Hill}, {Polyansky},
  {Mellau}, {Yurchenko}, \& {Tennyson}}]{barber2014}
{Barber}, R.~J., {Strange}, J.~K., {Hill}, C., {et~al.} 2014, \mnras, 437, 1828

\bibitem[{{Barclay} {et~al.}(2021){Barclay}, {Kostov}, {Col{\'o}n}, {Quintana},
  {Schlieder}, {Louie}, {Gilbert}, \& {Mullally}}]{2021AJ....162..300B}
{Barclay}, T., {Kostov}, V.~B., {Col{\'o}n}, K.~D., {et~al.} 2021, \aj, 162,
  300

\bibitem[{{Barclay} {et~al.}(2018){Barclay}, {Pepper}, \&
  {Quintana}}]{2018ApJS..239....2B}
{Barclay}, T., {Pepper}, J., \& {Quintana}, E.~V. 2018, \apjs, 239, 2

\bibitem[{{Benneke} \& {Seager}(2013)}]{Benneke2013}
{Benneke}, B., \& {Seager}, S. 2013, \apj, 778, 153

\bibitem[{{Buchner} {et~al.}(2014{\natexlab{a}}){Buchner}, {Georgakakis},
  {Nandra}, {Hsu}, {Rangel}, {Brightman}, {Merloni}, {Salvato}, {Donley}, \&
  {Kocevski}}]{Buchner2014}
{Buchner}, J., {Georgakakis}, A., {Nandra}, K., {et~al.} 2014{\natexlab{a}},
  \aap, 564, A125

\bibitem[{{Buchner} {et~al.}(2014{\natexlab{b}}){Buchner}, {Georgakakis},
  {Nandra}, {Hsu}, {Rangel}, {Brightman}, {Merloni}, {Salvato}, {Donley}, \&
  {Kocevski}}]{2014A&A...564A.125B}
---. 2014{\natexlab{b}}, \aap, 564, A125

\bibitem[{{Burt} {et~al.}(2021){Burt}, {Dragomir}, {Molli{\`e}re},
  {Youngblood}, {Garc{\'\i}a Mu{\~n}oz}, {McCann}, {Kreidberg}, {Huang},
  {Collins}, {Eastman}, {Abe}, {Almenara}, {Crossfield}, {Ziegler},
  {Rodriguez}, {Mamajek}, {Stassun}, {Halverson}, {Villanueva}, {Butler},
  {Wang}, {Schwarz}, {Ricker}, {Vanderspek}, {Latham}, {Seager}, {Winn},
  {Jenkins}, {Agabi}, {Bonfils}, {Ciardi}, {Cointepas}, {Crane}, {Crouzet},
  {Dransfield}, {Feng}, {Furlan}, {Guillot}, {Gupta}, {Howell}, {Jensen},
  {Law}, {Mann}, {Marie-Sainte}, {Matson}, {Matthews}, {M{\'e}karnia},
  {Pepper}, {Scott}, {Shectman}, {Schlieder}, {Schmider}, {Stevens}, {Teske},
  {Triaud}, {Charbonneau}, {Berta-Thompson}, {Burke}, {Daylan}, {Barclay},
  {Wohler}, \& {Brasseur}}]{2021AJ....162...87B}
{Burt}, J.~A., {Dragomir}, D., {Molli{\`e}re}, P., {et~al.} 2021, \aj, 162, 87

\bibitem[{{Castelli} \& {Kurucz}(2003)}]{2003IAUS..210P.A20C}
{Castelli}, F., \& {Kurucz}, R.~L. 2003, in Modelling of Stellar Atmospheres,
  ed. N.~{Piskunov}, W.~W. {Weiss}, \& D.~F. {Gray}, Vol. 210, A20

\bibitem[{{Charbonneau} {et~al.}(2009){Charbonneau}, {Berta}, {Irwin}, {Burke},
  {Nutzman}, {Buchhave}, {Lovis}, {Bonfils}, {Latham}, {Udry}, {Murray-Clay},
  {Holman}, {Falco}, {Winn}, {Queloz}, {Pepe}, {Mayor}, {Delfosse}, \&
  {Forveille}}]{2009Natur.462..891C}
{Charbonneau}, D., {Berta}, Z.~K., {Irwin}, J., {et~al.} 2009, \nat, 462, 891

\bibitem[{{Cloutier} {et~al.}(2020){Cloutier}, {Eastman}, {Rodriguez},
  {Astudillo-Defru}, {Bonfils}, {Mortier}, {Watson}, {Stalport}, {Pinamonti},
  {Lienhard}, {Harutyunyan}, {Damasso}, {Latham}, {Collins}, {Massey}, {Irwin},
  {Winters}, {Charbonneau}, {Ziegler}, {Matthews}, {Crossfield}, {Kreidberg},
  {Quinn}, {Ricker}, {Vanderspek}, {Seager}, {Winn}, {Jenkins}, {Vezie},
  {Udry}, {Twicken}, {Tenenbaum}, {Sozzetti}, {S{\'e}gransan}, {Schlieder},
  {Sasselov}, {Santos}, {Rice}, {Rackham}, {Poretti}, {Piotto}, {Phillips},
  {Pepe}, {Molinari}, {Mignon}, {Micela}, {Melo}, {de Medeiros}, {Mayor},
  {Matson}, {Martinez Fiorenzano}, {Mann}, {Magazz{\'u}}, {Lovis},
  {L{\'o}pez-Morales}, {Lopez}, {Lissauer}, {L{\'e}pine}, {Law}, {Kielkopf},
  {Johnson}, {Jensen}, {Howell}, {Gonzales}, {Ghedina}, {Forveille},
  {Figueira}, {Dumusque}, {Dressing}, {Doyon}, {D{\'\i}az}, {Fabrizio},
  {Delfosse}, {Cosentino}, {Conti}, {Collins}, {Cameron}, {Ciardi}, {Caldwell},
  {Burke}, {Buchhave}, {Brice{\~n}o}, {Boyd}, {Bouchy}, {Beichman}, {Artigau},
  \& {Almenara}}]{2020AJ....160....3C}
{Cloutier}, R., {Eastman}, J.~D., {Rodriguez}, J.~E., {et~al.} 2020, \aj, 160,
  3

\bibitem[{{Cloutier} {et~al.}(2021){Cloutier}, {Charbonneau}, {Stassun},
  {Murgas}, {Mortier}, {Massey}, {Lissauer}, {Latham}, {Irwin}, {Haywood},
  {Guerra}, {Girardin}, {Giacalone}, {Bosch-Cabot}, {Bieryla}, {Winn},
  {Watson}, {Vanderspek}, {Udry}, {Tamura}, {Sozzetti}, {Shporer},
  {S{\'e}gransan}, {Seager}, {Savel}, {Sasselov}, {Rose}, {Ricker}, {Rice},
  {Quintana}, {Quinn}, {Piotto}, {Phillips}, {Pepe}, {Pedani}, {Parviainen},
  {Palle}, {Narita}, {Molinari}, {Micela}, {McDermott}, {Mayor}, {Matson},
  {Martinez Fiorenzano}, {Lovis}, {L{\'o}pez-Morales}, {Kusakabe}, {Jensen},
  {Jenkins}, {Huang}, {Howell}, {Harutyunyan}, {F{\H{u}}r{\'e}sz}, {Fukui},
  {Esquerdo}, {Esparza-Borges}, {Dumusque}, {Dressing}, {Fabrizio}, {Collins},
  {Cameron}, {Christiansen}, {Cecconi}, {Buchhave}, {Boschin}, \&
  {Andreuzzi}}]{2021AJ....162...79C}
{Cloutier}, R., {Charbonneau}, D., {Stassun}, K.~G., {et~al.} 2021, \aj, 162,
  79

\bibitem[{{Constantinou} \& {Madhusudhan}(2022)}]{Constantinou2022}
{Constantinou}, S., \& {Madhusudhan}, N. 2022, \mnras, 514, 2073

\bibitem[{{Cram} \& {Mullan}(1985)}]{Cram1985}
{Cram}, L.~E., \& {Mullan}, D.~J. 1985, \apj, 294, 626

\bibitem[{{Crossfield} {et~al.}(2015){Crossfield}, {Petigura}, {Schlieder},
  {Howard}, {Fulton}, {Aller}, {Ciardi}, {L{\'e}pine}, {Barclay}, {de Pater},
  {de Kleer}, {Quintana}, {Christiansen}, {Schlafly}, {Kaltenegger}, {Crepp},
  {Henning}, {Obermeier}, {Deacon}, {Weiss}, {Isaacson}, {Hansen}, {Liu},
  {Greene}, {Howell}, {Barman}, \& {Mordasini}}]{2015ApJ...804...10C}
{Crossfield}, I. J.~M., {Petigura}, E., {Schlieder}, J.~E., {et~al.} 2015,
  \apj, 804, 10

\bibitem[{{Deming} {et~al.}(2013){Deming}, {Wilkins}, {McCullough}, {Burrows},
  {Fortney}, {Agol}, {Dobbs-Dixon}, {Madhusudhan}, {Crouzet}, {Desert},
  {Gilliland}, {Haynes}, {Knutson}, {Line}, {Magic}, {Mandell}, {Ranjan},
  {Charbonneau}, {Clampin}, {Seager}, \& {Showman}}]{2013ApJ...774...95D}
{Deming}, D., {Wilkins}, A., {McCullough}, P., {et~al.} 2013, \apj, 774, 95

\bibitem[{{Demory} {et~al.}(2020){Demory}, {Pozuelos}, {G{\'o}mez Maqueo Chew},
  {Sabin}, {Petrucci}, {Schroffenegger}, {Grimm}, {Sestovic}, {Gillon},
  {McCormac}, {Barkaoui}, {Benz}, {Bieryla}, {Bouchy}, {Burdanov}, {Collins},
  {de Wit}, {Dressing}, {Garcia}, {Giacalone}, {Guerra}, {Haldemann}, {Heng},
  {Jehin}, {Jofr{\'e}}, {Kane}, {Lillo-Box}, {Maign{\'e}}, {Mordasini},
  {Morris}, {Niraula}, {Queloz}, {Rackham}, {Savel}, {Soubkiou}, {Srdoc},
  {Stassun}, {Triaud}, {Zambelli}, {Ricker}, {Latham}, {Seager}, {Winn},
  {Jenkins}, {Calvario-Vel{\'a}squez}, {Franco Herrera}, {Colorado}, {Cadena
  Zepeda}, {Figueroa}, {Watson}, {Lugo-Ibarra}, {Carigi}, {Guisa}, {Herrera},
  {Sierra D{\'\i}az}, {Su{\'a}rez}, {Barrado}, {Batalha}, {Benkhaldoun},
  {Chontos}, {Dai}, {Essack}, {Ghachoui}, {Huang}, {Huber}, {Isaacson},
  {Lissauer}, {Morales-Calder{\'o}n}, {Robertson}, {Roy}, {Twicken},
  {Vanderburg}, \& {Weiss}}]{2020A&A...642A..49D}
{Demory}, B.~O., {Pozuelos}, F.~J., {G{\'o}mez Maqueo Chew}, Y., {et~al.} 2020,
  \aap, 642, A49

\bibitem[{{Evans} {et~al.}(2016){Evans}, {Sing}, {Wakeford}, {Nikolov},
  {Ballester}, {Drummond}, {Kataria}, {Gibson}, {Amundsen}, \&
  {Spake}}]{2016ApJ...822L...4E}
{Evans}, T.~M., {Sing}, D.~K., {Wakeford}, H.~R., {et~al.} 2016, \apjl, 822, L4

\bibitem[{{Feroz} {et~al.}(2009){Feroz}, {Hobson}, \& {Bridges}}]{Feroz2009}
{Feroz}, F., {Hobson}, M.~P., \& {Bridges}, M. 2009, \mnras, 398, 1601

\bibitem[{{Figueira} {et~al.}(2009){Figueira}, {Pont}, {Mordasini}, {Alibert},
  {Georgy}, \& {Benz}}]{2009A&A...493..671F}
{Figueira}, P., {Pont}, F., {Mordasini}, C., {et~al.} 2009, \aap, 493, 671

\bibitem[{{Foreman-Mackey} {et~al.}(2013){Foreman-Mackey}, {Hogg}, {Lang}, \&
  {Goodman}}]{2013PASP..125..306F}
{Foreman-Mackey}, D., {Hogg}, D.~W., {Lang}, D., \& {Goodman}, J. 2013, \pasp,
  125, 306

\bibitem[{{Fortney} {et~al.}(2013){Fortney}, {Mordasini}, {Nettelmann},
  {Kempton}, {Greene}, \& {Zahnle}}]{Fortney2013}
{Fortney}, J.~J., {Mordasini}, C., {Nettelmann}, N., {et~al.} 2013, \apj, 775,
  80

\bibitem[{{Fulton} {et~al.}(2017){Fulton}, {Petigura}, {Howard}, {Isaacson},
  {Marcy}, {Cargile}, {Hebb}, {Weiss}, {Johnson}, {Morton}, {Sinukoff},
  {Crossfield}, \& {Hirsch}}]{2017AJ....154..109F}
{Fulton}, B.~J., {Petigura}, E.~A., {Howard}, A.~W., {et~al.} 2017, \aj, 154,
  109

\bibitem[{{Gandhi} \& {Madhusudhan}(2017)}]{gandhi2017}
{Gandhi}, S., \& {Madhusudhan}, N. 2017, \mnras, 472, 2334

\bibitem[{{Gillon} {et~al.}(2007){Gillon}, {Pont}, {Demory}, {Mallmann},
  {Mayor}, {Mazeh}, {Queloz}, {Shporer}, {Udry}, \&
  {Vuissoz}}]{2007A&A...472L..13G}
{Gillon}, M., {Pont}, F., {Demory}, B.~O., {et~al.} 2007, \aap, 472, L13

\bibitem[{{Guerrero} {et~al.}(2021){Guerrero}, {Seager}, {Huang}, {Vanderburg},
  {Garcia Soto}, {Mireles}, {Hesse}, {Fong}, {Glidden}, {Shporer}, {Latham},
  {Collins}, {Quinn}, {Burt}, {Dragomir}, {Crossfield}, {Vanderspek},
  {Fausnaugh}, {Burke}, {Ricker}, {Daylan}, {Essack}, {G{\"u}nther}, {Osborn},
  {Pepper}, {Rowden}, {Sha}, {Villanueva}, {Yahalomi}, {Yu}, {Ballard},
  {Batalha}, {Berardo}, {Chontos}, {Dittmann}, {Esquerdo}, {Mikal-Evans},
  {Jayaraman}, {Krishnamurthy}, {Louie}, {Mehrle}, {Niraula}, {Rackham},
  {Rodriguez}, {Rowden}, {Sousa-Silva}, {Watanabe}, {Wong}, {Zhan},
  {Zivanovic}, {Christiansen}, {Ciardi}, {Swain}, {Lund}, {Mullally},
  {Fleming}, {Rodriguez}, {Boyd}, {Quintana}, {Barclay}, {Col{\'o}n},
  {Rinehart}, {Schlieder}, {Clampin}, {Jenkins}, {Twicken}, {Caldwell},
  {Coughlin}, {Henze}, {Lissauer}, {Morris}, {Rose}, {Smith}, {Tenenbaum},
  {Ting}, {Wohler}, {Bakos}, {Bean}, {Berta-Thompson}, {Bieryla}, {Bouma},
  {Buchhave}, {Butler}, {Charbonneau}, {Doty}, {Ge}, {Holman}, {Howard},
  {Kaltenegger}, {Kane}, {Kjeldsen}, {Kreidberg}, {Lin}, {Minsky}, {Narita},
  {Paegert}, {P{\'a}l}, {Palle}, {Sasselov}, {Spencer}, {Sozzetti}, {Stassun},
  {Torres}, {Udry}, \& {Winn}}]{2021ApJS..254...39G}
{Guerrero}, N.~M., {Seager}, S., {Huang}, C.~X., {et~al.} 2021, \apjs, 254, 39

\bibitem[{{G{\"u}nther} {et~al.}(2019){G{\"u}nther}, {Pozuelos}, {Dittmann},
  {Dragomir}, {Kane}, {Daylan}, {Feinstein}, {Huang}, {Morton}, {Bonfanti},
  {Bouma}, {Burt}, {Collins}, {Lissauer}, {Matthews}, {Montet}, {Vanderburg},
  {Wang}, {Winters}, {Ricker}, {Vanderspek}, {Latham}, {Seager}, {Winn},
  {Jenkins}, {Armstrong}, {Barkaoui}, {Batalha}, {Bean}, {Caldwell}, {Ciardi},
  {Collins}, {Crossfield}, {Fausnaugh}, {Furesz}, {Gan}, {Gillon}, {Guerrero},
  {Horne}, {Howell}, {Ireland}, {Isopi}, {Jehin}, {Kielkopf}, {Lepine},
  {Mallia}, {Matson}, {Myers}, {Palle}, {Quinn}, {Relles}, {Rojas-Ayala},
  {Schlieder}, {Sefako}, {Shporer}, {Su{\'a}rez}, {Tan}, {Ting}, {Twicken}, \&
  {Waite}}]{Guenther2019}
{G{\"u}nther}, M.~N., {Pozuelos}, F.~J., {Dittmann}, J.~A., {et~al.} 2019,
  Nature Astronomy, 3, 1099

\bibitem[{{Hirano} {et~al.}(2021){Hirano}, {Livingston}, {Fukui}, {Narita},
  {Harakawa}, {Ishikawa}, {Miyakawa}, {Kimura}, {Nakayama}, {Fujita}, {Hori},
  {Stassun}, {Bieryla}, {Cadieux}, {Ciardi}, {Collins}, {Ikoma}, {Vanderburg},
  {Barclay}, {Brasseur}, {de Leon}, {Doty}, {Doyon}, {Esparza-Borges},
  {Esquerdo}, {Furlan}, {Gaidos}, {Gonzales}, {Hodapp}, {Howell}, {Isogai},
  {Jacobson}, {Jenkins}, {Jensen}, {Kawauchi}, {Kotani}, {Kudo}, {Kurita},
  {Kurokawa}, {Kusakabe}, {Kuzuhara}, {Lafreni{\`e}re}, {Latham}, {Massey},
  {Mori}, {Murgas}, {Nishikawa}, {Nishiumi}, {Omiya}, {Paegert}, {Palle},
  {Parviainen}, {Quinn}, {Ricker}, {Schwarz}, {Seager}, {Tamura}, {Tenenbaum},
  {Terada}, {Vanderspek}, {Vievard}, {Watanabe}, \&
  {Winn}}]{2021AJ....162..161H}
{Hirano}, T., {Livingston}, J.~H., {Fukui}, A., {et~al.} 2021, \aj, 162, 161

\bibitem[{{Hu} {et~al.}(2021){Hu}, {Damiano}, {Scheucher}, {Kite}, {Seager}, \&
  {Rauer}}]{Hu2021}
{Hu}, R., {Damiano}, M., {Scheucher}, M., {et~al.} 2021, \apjl, 921, L8

\bibitem[{{Hunter}(2007)}]{2007CSE.....9...90H}
{Hunter}, J.~D. 2007, Computing in Science and Engineering, 9, 90

\bibitem[{Husser {et~al.}(2013)Husser, {Wende-von Berg}, Dreizler, Homeier,
  Reiners, Barman, \& Hauschildt}]{Husser2013}
Husser, T.-O., {Wende-von Berg}, S., Dreizler, S., {et~al.} 2013, A{\&}A, 553,
  A6

\bibitem[{{Kaye} {et~al.}(2022){Kaye}, {Vissapragada}, {G{\"u}nther},
  {Aigrain}, {Mikal-Evans}, {Jensen}, {Parviainen}, {Pozuelos}, {Abe}, {Acton},
  {Agabi}, {Alves}, {Anderson}, {Armstrong}, {Barkaoui}, {Barrag{\'a}n},
  {Benneke}, {Boyd}, {Brahm}, {Bruni}, {Bryant}, {Burleigh}, {Casewell},
  {Ciardi}, {Cloutier}, {Collins}, {Collins}, {Conti}, {Crossfield}, {Crouzet},
  {Daylan}, {Dragomir}, {Dransfield}, {Fabrycky}, {Fausnaugh},
  {Fu{\H{u}}r{\'e}sz}, {Gan}, {Gill}, {Gillon}, {Goad}, {Gorjian},
  {Greklek-McKeon}, {Guerrero}, {Guillot}, {Jehin}, {Jenkins}, {Lendl},
  {Kamler}, {Kane}, {Kielkopf}, {Kunimoto}, {Marie-Sainte}, {McCormac},
  {M{\'e}karnia}, {Morales}, {Moyano}, {Palle}, {Parmentier}, {Relles},
  {Schmider}, {Schwarz}, {Seager}, {Smith}, {Tan}, {Taylor}, {Triaud},
  {Twicken}, {Udry}, {Vines}, {Wang}, {Wheatley}, \& {Winn}}]{Kaye2022}
{Kaye}, L., {Vissapragada}, S., {G{\"u}nther}, M.~N., {et~al.} 2022, \mnras,
  510, 5464

\bibitem[{{Kossakowski} {et~al.}(2021){Kossakowski}, {Kemmer}, {Bluhm},
  {Stock}, {Caballero}, {B{\'e}jar}, {Guill{\'e}n}, {Lodieu}, {Collins},
  {Oshagh}, {Schlecker}, {Espinoza}, {Pall{\'e}}, {Henning}, {Kreidberg},
  {K{\"u}rster}, {Amado}, {Anderson}, {Morales}, {Cartwright}, {Charbonneau},
  {Chaturvedi}, {Cifuentes}, {Conti}, {Cort{\'e}s-Contreras}, {Dreizler},
  {Galad{\'\i}-Enr{\'\i}quez}, {Guerra}, {Hart}, {Hellier}, {Henze}, {Herrero},
  {Jeffers}, {Jenkins}, {Jensen}, {Kaminski}, {Kielkopf}, {Kunimoto},
  {Lafarga}, {Latham}, {Lillo-Box}, {Luque}, {Molaverdikhani}, {Montes},
  {Morello}, {Morgan}, {Nowak}, {Pavlov}, {Perger}, {Quintana}, {Quirrenbach},
  {Reffert}, {Reiners}, {Ricker}, {Ribas}, {L{\'o}pez}, {Osorio}, {Seager},
  {Sch{\"o}fer}, {Schweitzer}, {Trifonov}, {Vanaverbeke}, {Vanderspek}, {West},
  {Winn}, \& {Zechmeister}}]{2021A&A...656A.124K}
{Kossakowski}, D., {Kemmer}, J., {Bluhm}, P., {et~al.} 2021, \aap, 656, A124

\bibitem[{{Kreidberg}(2015)}]{2015PASP..127.1161K}
{Kreidberg}, L. 2015, \pasp, 127, 1161

\bibitem[{{Kunimoto} {et~al.}(2022){Kunimoto}, {Winn}, {Ricker}, \&
  {Vanderspek}}]{2022AJ....163..290K}
{Kunimoto}, M., {Winn}, J., {Ricker}, G.~R., \& {Vanderspek}, R.~K. 2022, \aj,
  163, 290

\bibitem[{{Kurucz}(1993)}]{1993KurCD..13.....K}
{Kurucz}, R. 1993, ATLAS9 Stellar Atmosphere Programs and 2 km/s grid.~Kurucz
  CD-ROM No.~13.~ Cambridge, Mass.: Smithsonian Astrophysical Observatory

\bibitem[{{Luque} {et~al.}(2021){Luque}, {Serrano}, {Molaverdikhani}, {Nixon},
  {Livingston}, {Guenther}, {Pall{\'e}}, {Madhusudhan}, {Nowak}, {Korth},
  {Cochran}, {Hirano}, {Chaturvedi}, {Goffo}, {Albrecht}, {Barrag{\'a}n},
  {Brice{\~n}o}, {Cabrera}, {Charbonneau}, {Cloutier}, {Collins}, {Collins},
  {Col{\'o}n}, {Crossfield}, {Csizmadia}, {Dai}, {Deeg}, {Esposito},
  {Fridlund}, {Gandolfi}, {Georgieva}, {Glidden}, {Goeke}, {Grziwa}, {Hatzes},
  {Henze}, {Howell}, {Irwin}, {Jenkins}, {Jensen}, {K{\'a}bath}, {Kidwell},
  {Kielkopf}, {Knudstrup}, {Lam}, {Latham}, {Lissauer}, {Mann}, {Matthews},
  {Mireles}, {Narita}, {Paegert}, {Persson}, {Redfield}, {Ricker}, {Rodler},
  {Schlieder}, {Scott}, {Seager}, {{\v{S}}ubjak}, {Tan}, {Ting}, {Vanderspek},
  {Van Eylen}, {Winn}, \& {Ziegler}}]{2021A&A...645A..41L}
{Luque}, R., {Serrano}, L.~M., {Molaverdikhani}, K., {et~al.} 2021, \aap, 645,
  A41

\bibitem[{{Madhusudhan} {et~al.}(2020){Madhusudhan}, {Nixon}, {Welbanks},
  {Piette}, \& {Booth}}]{Madhu2020}
{Madhusudhan}, N., {Nixon}, M.~C., {Welbanks}, L., {Piette}, A. A.~A., \&
  {Booth}, R.~A. 2020, \apjl, 891, L7

\bibitem[{{Madhusudhan} {et~al.}(2021){Madhusudhan}, {Piette}, \&
  {Constantinou}}]{2021ApJ...918....1M}
{Madhusudhan}, N., {Piette}, A. A.~A., \& {Constantinou}, S. 2021, \apj, 918, 1

\bibitem[{{Madhusudhan} \& {Seager}(2011)}]{Madhu2011}
{Madhusudhan}, N., \& {Seager}, S. 2011, \apj, 729, 41

\bibitem[{{Martioli} {et~al.}(2021){Martioli}, {H{\'e}brard}, {Correia},
  {Laskar}, \& {Lecavelier des Etangs}}]{2021A&A...649A.177M}
{Martioli}, E., {H{\'e}brard}, G., {Correia}, A.~C.~M., {Laskar}, J., \&
  {Lecavelier des Etangs}, A. 2021, \aap, 649, A177

\bibitem[{{McCullough} {et~al.}(2014){McCullough}, {Crouzet}, {Deming}, \&
  {Madhusudhan}}]{2014ApJ...791...55M}
{McCullough}, P.~R., {Crouzet}, N., {Deming}, D., \& {Madhusudhan}, N. 2014,
  \apj, 791, 55

\bibitem[{{Mikal-Evans} {et~al.}(2020){Mikal-Evans}, {Sing}, {Kataria},
  {Wakeford}, {Mayne}, {Lewis}, {Barstow}, \& {Spake}}]{2020MNRAS.496.1638M}
{Mikal-Evans}, T., {Sing}, D.~K., {Kataria}, T., {et~al.} 2020, \mnras, 496,
  1638

\bibitem[{{Mikal-Evans} {et~al.}(2019{\natexlab{a}}){Mikal-Evans}, {Sing},
  {Goyal}, {Drummond}, {Carter}, {Henry}, {Wakeford}, {Lewis}, {Marley},
  {Tremblin}, {Nikolov}, {Kataria}, {Deming}, \&
  {Ballester}}]{2019MNRAS.488.2222M}
{Mikal-Evans}, T., {Sing}, D.~K., {Goyal}, J.~M., {et~al.} 2019{\natexlab{a}},
  \mnras, 488, 2222

\bibitem[{{Mikal-Evans} {et~al.}(2019{\natexlab{b}}){Mikal-Evans},
  {Crossfield}, {Daylan}, {Delrez}, {Dittmann}, {Drummond}, {Guenther},
  {Kreidberg}, {Madhusudhan}, {Triaud}, {Van Eylen}, \&
  {Welbanks}}]{2019hst..prop15814M}
{Mikal-Evans}, T., {Crossfield}, I., {Daylan}, T., {et~al.} 2019{\natexlab{b}},
  {Atmospheric characterization of two temperate mini-Neptunes formed in the
  same protoplanetary nebula}, HST Proposal

\bibitem[{{Mikal-Evans} {et~al.}(2021){Mikal-Evans}, {Crossfield}, {Benneke},
  {Kreidberg}, {Moses}, {Morley}, {Thorngren}, {Molli{\`e}re},
  {Hardegree-Ullman}, {Brewer}, {Christiansen}, {Ciardi}, {Dragomir},
  {Dressing}, {Fortney}, {Gorjian}, {Greene}, {Hirsch}, {Howard}, {Howell},
  {Isaacson}, {Kosiarek}, {Krick}, {Livingston}, {Lothringer}, {Morales},
  {Petigura}, {Schlieder}, \& {Werner}}]{2021AJ....161...18M}
{Mikal-Evans}, T., {Crossfield}, I. J.~M., {Benneke}, B., {et~al.} 2021, \aj,
  161, 18

\bibitem[{{Mikal-Evans} {et~al.}(2022){Mikal-Evans}, {Sing}, {Barstow},
  {Kataria}, {Goyal}, {Lewis}, {Taylor}, {Mayne}, {Daylan}, {Wakeford},
  {Marley}, \& {Spake}}]{2022NatAs...6..471M}
{Mikal-Evans}, T., {Sing}, D.~K., {Barstow}, J.~K., {et~al.} 2022, Nature
  Astronomy, 6, 471

\bibitem[{{Montet} {et~al.}(2015){Montet}, {Morton}, {Foreman-Mackey},
  {Johnson}, {Hogg}, {Bowler}, {Latham}, {Bieryla}, \&
  {Mann}}]{2015ApJ...809...25M}
{Montet}, B.~T., {Morton}, T.~D., {Foreman-Mackey}, D., {et~al.} 2015, \apj,
  809, 25

\bibitem[{{Moses} {et~al.}(2013){Moses}, {Line}, {Visscher}, {Richardson},
  {Nettelmann}, {Fortney}, {Barman}, {Stevenson}, \& {Madhusudhan}}]{Moses2013}
{Moses}, J.~I., {Line}, M.~R., {Visscher}, C., {et~al.} 2013, \apj, 777, 34

\bibitem[{{Newton} {et~al.}(2016){Newton}, {Irwin}, {Charbonneau},
  {Berta-Thompson}, {Dittmann}, \& {West}}]{Newton2016}
{Newton}, E.~R., {Irwin}, J., {Charbonneau}, D., {et~al.} 2016, \apj, 821, 93

\bibitem[{{Nowak} {et~al.}(2020){Nowak}, {Luque}, {Parviainen}, {Pall{\'e}},
  {Molaverdikhani}, {B{\'e}jar}, {Lillo-Box}, {Rodr{\'\i}guez-L{\'o}pez},
  {Caballero}, {Zechmeister}, {Passegger}, {Cifuentes}, {Schweitzer}, {Narita},
  {Cale}, {Espinoza}, {Murgas}, {Hidalgo}, {Zapatero Osorio}, {Pozuelos},
  {Aceituno}, {Amado}, {Barkaoui}, {Barrado}, {Bauer}, {Benkhaldoun},
  {Caldwell}, {Casasayas Barris}, {Chaturvedi}, {Chen}, {Collins}, {Collins},
  {Cort{\'e}s-Contreras}, {Crossfield}, {de Le{\'o}n}, {D{\'\i}ez Alonso},
  {Dreizler}, {El Mufti}, {Esparza-Borges}, {Essack}, {Fukui}, {Gaidos},
  {Gillon}, {Gonzales}, {Guerra}, {Hatzes}, {Henning}, {Herrero}, {Hesse},
  {Hirano}, {Howell}, {Jeffers}, {Jehin}, {Jenkins}, {Kaminski}, {Kemmer},
  {Kielkopf}, {Kossakowski}, {Kotani}, {K{\"u}rster}, {Lafarga}, {Latham},
  {Law}, {Lissauer}, {Lodieu}, {Madrigal-Aguado}, {Mann}, {Massey}, {Matson},
  {Matthews}, {Monta{\~n}{\'e}s-Rodr{\'\i}guez}, {Montes}, {Morales}, {Mori},
  {Nagel}, {Oshagh}, {Pedraz}, {Plavchan}, {Pollacco}, {Quirrenbach},
  {Reffert}, {Reiners}, {Ribas}, {Ricker}, {Rose}, {Schlecker}, {Schlieder},
  {Seager}, {Stangret}, {Stock}, {Tamura}, {Tanner}, {Teske}, {Trifonov},
  {Twicken}, {Vanderspek}, {Watanabe}, {Wittrock}, {Ziegler}, \&
  {Zohrabi}}]{2020A&A...642A.173N}
{Nowak}, G., {Luque}, R., {Parviainen}, H., {et~al.} 2020, \aap, 642, A173

\bibitem[{Parviainen \& Aigrain(2015)}]{Parviainen2015}
Parviainen, H., \& Aigrain, S. 2015, MNRAS, 453, 3821

\bibitem[{{Petigura} {et~al.}(2022){Petigura}, {Rogers}, {Isaacson}, {Owen},
  {Kraus}, {Winn}, {MacDougall}, {Howard}, {Fulton}, {Kosiarek}, {Weiss},
  {Behmard}, \& {Blunt}}]{2022AJ....163..179P}
{Petigura}, E.~A., {Rogers}, J.~G., {Isaacson}, H., {et~al.} 2022, \aj, 163,
  179

\bibitem[{{Pinhas} {et~al.}(2018){Pinhas}, {Rackham}, {Madhusudhan}, \&
  {Apai}}]{pinhas2018}
{Pinhas}, A., {Rackham}, B.~V., {Madhusudhan}, N., \& {Apai}, D. 2018, \mnras,
  480, 5314

\bibitem[{{Plavchan} {et~al.}(2020){Plavchan}, {Barclay}, {Gagn{\'e}}, {Gao},
  {Cale}, {Matzko}, {Dragomir}, {Quinn}, {Feliz}, {Stassun}, {Crossfield},
  {Berardo}, {Latham}, {Tieu}, {Anglada-Escud{\'e}}, {Ricker}, {Vanderspek},
  {Seager}, {Winn}, {Jenkins}, {Rinehart}, {Krishnamurthy}, {Dynes}, {Doty},
  {Adams}, {Afanasev}, {Beichman}, {Bottom}, {Bowler}, {Brinkworth}, {Brown},
  {Cancino}, {Ciardi}, {Clampin}, {Clark}, {Collins}, {Davison},
  {Foreman-Mackey}, {Furlan}, {Gaidos}, {Geneser}, {Giddens}, {Gilbert},
  {Hall}, {Hellier}, {Henry}, {Horner}, {Howard}, {Huang}, {Huber}, {Kane},
  {Kenworthy}, {Kielkopf}, {Kipping}, {Klenke}, {Kruse}, {Latouf}, {Lowrance},
  {Mennesson}, {Mengel}, {Mills}, {Morton}, {Narita}, {Newton}, {Nishimoto},
  {Okumura}, {Palle}, {Pepper}, {Quintana}, {Roberge}, {Roccatagliata},
  {Schlieder}, {Tanner}, {Teske}, {Tinney}, {Vanderburg}, {von Braun}, {Walp},
  {Wang}, {Wang}, {Weigand}, {White}, {Wittenmyer}, {Wright}, {Youngblood},
  {Zhang}, \& {Zilberman}}]{2020Natur.582..497P}
{Plavchan}, P., {Barclay}, T., {Gagn{\'e}}, J., {et~al.} 2020, \nat, 582, 497

\bibitem[{{Rackham} {et~al.}(2018){Rackham}, {Apai}, \&
  {Giampapa}}]{rackham2018}
{Rackham}, B.~V., {Apai}, D., \& {Giampapa}, M.~S. 2018, \apj, 853, 122

\bibitem[{{Richard} {et~al.}(2012){Richard}, {Gordon}, {Rothman}, {Abel},
  {Frommhold}, {Gustafsson}, {Hartmann}, {Hermans}, {Lafferty}, {Orton},
  {Smith}, \& {Tran}}]{richard2012}
{Richard}, C., {Gordon}, I.~E., {Rothman}, L.~S., {et~al.} 2012, \jqsrt, 113,
  1276

\bibitem[{{Ricker} {et~al.}(2015){Ricker}, {Winn}, {Vanderspek}, {Latham},
  {Bakos}, {Bean}, {Berta-Thompson}, {Brown}, {Buchhave}, {Butler}, {Butler},
  {Chaplin}, {Charbonneau}, {Christensen-Dalsgaard}, {Clampin}, {Deming},
  {Doty}, {De Lee}, {Dressing}, {Dunham}, {Endl}, {Fressin}, {Ge}, {Henning},
  {Holman}, {Howard}, {Ida}, {Jenkins}, {Jernigan}, {Johnson}, {Kaltenegger},
  {Kawai}, {Kjeldsen}, {Laughlin}, {Levine}, {Lin}, {Lissauer}, {MacQueen},
  {Marcy}, {McCullough}, {Morton}, {Narita}, {Paegert}, {Palle}, {Pepe},
  {Pepper}, {Quirrenbach}, {Rinehart}, {Sasselov}, {Sato}, {Seager},
  {Sozzetti}, {Stassun}, {Sullivan}, {Szentgyorgyi}, {Torres}, {Udry}, \&
  {Villasenor}}]{2015JATIS...1a4003R}
{Ricker}, G.~R., {Winn}, J.~N., {Vanderspek}, R., {et~al.} 2015, Journal of
  Astronomical Telescopes, Instruments, and Systems, 1, 014003

\bibitem[{Robertson {et~al.}(2013)Robertson, Endl, Cochran, \&
  Dodson-Robinson}]{Robertson2013}
Robertson, P., Endl, M., Cochran, W.~D., \& Dodson-Robinson, S.~E. 2013, The
  Astrophysical Journal, 764, 3

\bibitem[{{Rogers} \& {Seager}(2010)}]{2010ApJ...716.1208R}
{Rogers}, L.~A., \& {Seager}, S. 2010, \apj, 716, 1208

\bibitem[{{Rothman} {et~al.}(2010){Rothman}, {Gordon}, {Barber}, {Dothe},
  {Gamache}, {Goldman}, {Perevalov}, {Tashkun}, \& {Tennyson}}]{rothman2010}
{Rothman}, L.~S., {Gordon}, I.~E., {Barber}, R.~J., {et~al.} 2010, \jqsrt, 111,
  2139

\bibitem[{{Stauffer} \& {Hartmann}(1986)}]{Stauffer1986}
{Stauffer}, J.~R., \& {Hartmann}, L.~W. 1986, \apjs, 61, 531

\bibitem[{{Stef{\'a}nsson} {et~al.}(2020){Stef{\'a}nsson}, {Kopparapu}, {Lin},
  {Mahadevan}, {Ca{\~n}as}, {Kanodia}, {Ninan}, {Cochran}, {Endl}, {Hebb},
  {Wisniewski}, {Gupta}, {Everett}, {Bender}, {Diddams}, {Ford}, {Fredrick},
  {Halverson}, {Hearty}, {Levi}, {Maney}, {Metcalf}, {Monson}, {Ramsey},
  {Robertson}, {Roy}, {Schwab}, {Terrien}, \& {Wright}}]{2020AJ....160..259S}
{Stef{\'a}nsson}, G., {Kopparapu}, R., {Lin}, A., {et~al.} 2020, \aj, 160, 259

\bibitem[{{STScI Development Team}(2013)}]{2013ascl.soft03023S}
{STScI Development Team}. 2013, {pysynphot: Synthetic photometry software
  package}, ascl:1303.023

\bibitem[{{Sullivan} {et~al.}(2015){Sullivan}, {Winn}, {Berta-Thompson},
  {Charbonneau}, {Deming}, {Dressing}, {Latham}, {Levine}, {McCullough},
  {Morton}, {Ricker}, {Vanderspek}, \& {Woods}}]{2015ApJ...809...77S}
{Sullivan}, P.~W., {Winn}, J.~N., {Berta-Thompson}, Z.~K., {et~al.} 2015, \apj,
  809, 77

\bibitem[{{Tsai} {et~al.}(2021){Tsai}, {Innes}, {Lichtenberg}, {Taylor},
  {Malik}, {Chubb}, \& {Pierrehumbert}}]{Tsai2021}
{Tsai}, S.-M., {Innes}, H., {Lichtenberg}, T., {et~al.} 2021, \apjl, 922, L27

\bibitem[{{van der Walt} {et~al.}(2011){van der Walt}, {Colbert}, \&
  {Varoquaux}}]{numpy2011}
{van der Walt}, S., {Colbert}, S.~C., \& {Varoquaux}, G. 2011, Computing in
  Science Engineering, 13, 22

\bibitem[{{Van Eylen} {et~al.}(2018){Van Eylen}, {Agentoft}, {Lundkvist},
  {Kjeldsen}, {Owen}, {Fulton}, {Petigura}, \& {Snellen}}]{2018MNRAS.479.4786V}
{Van Eylen}, V., {Agentoft}, C., {Lundkvist}, M.~S., {et~al.} 2018, \mnras,
  479, 4786

\bibitem[{{Van Eylen} {et~al.}(2021){Van Eylen}, {Astudillo-Defru}, {Bonfils},
  {Livingston}, {Hirano}, {Luque}, {Lam}, {Justesen}, {Winn}, {Gandolfi},
  {Nowak}, {Palle}, {Albrecht}, {Dai}, {Campos Estrada}, {Owen},
  {Foreman-Mackey}, {Fridlund}, {Korth}, {Mathur}, {Forveille}, {Mikal-Evans},
  {Osborne}, {Ho}, {Almenara}, {Artigau}, {Barrag{\'a}n}, {Barros}, {Bouchy},
  {Cabrera}, {Caldwell}, {Charbonneau}, {Chaturvedi}, {Cochran}, {Csizmadia},
  {Damasso}, {Delfosse}, {De Medeiros}, {D{\'\i}az}, {Doyon}, {Esposito},
  {F{\H{u}}r{\'e}sz}, {Figueira}, {Georgieva}, {Goffo}, {Grziwa}, {Guenther},
  {Hatzes}, {Jenkins}, {Kabath}, {Knudstrup}, {Latham}, {Lavie}, {Lovis},
  {Mennickent}, {Mullally}, {Murgas}, {Narita}, {Pepe}, {Persson}, {Redfield},
  {Ricker}, {Santos}, {Seager}, {Serrano}, {Smith}, {Su{\'a}rez Mascare{\~n}o},
  {Subjak}, {Twicken}, {Udry}, {Vanderspek}, \& {Zapatero
  Osorio}}]{VanEylen2021}
{Van Eylen}, V., {Astudillo-Defru}, N., {Bonfils}, X., {et~al.} 2021, \mnras,
  507, 2154

\bibitem[{{Virtanen} {et~al.}(2020){Virtanen}, {Gommers}, {Oliphant},
  {Haberland}, {Reddy}, {Cournapeau}, {Burovski}, {Peterson}, {Weckesser},
  {Bright}, {van der Walt}, {Brett}, {Wilson}, {Jarrod Millman}, {Mayorov},
  {Nelson}, {Jones}, {Kern}, {Larson}, {Carey}, {Polat}, {Feng}, {Moore}, {Vand
  erPlas}, {Laxalde}, {Perktold}, {Cimrman}, {Henriksen}, {Quintero}, {Harris},
  {Archibald}, {Ribeiro}, {Pedregosa}, {van Mulbregt}, \&
  {Contributors}}]{2020SciPy-NMeth}
{Virtanen}, P., {Gommers}, R., {Oliphant}, T.~E., {et~al.} 2020, Nature
  Methods, 17, 261

\bibitem[{{Welbanks} \& {Madhusudhan}(2021)}]{Welbanks2021}
{Welbanks}, L., \& {Madhusudhan}, N. 2021, arXiv e-prints, arXiv:2103.08600

\bibitem[{{Welbanks} {et~al.}(2019){Welbanks}, {Madhusudhan}, {Allard},
  {Hubeny}, {Spiegelman}, \& {Leininger}}]{Welbanks2019}
{Welbanks}, L., {Madhusudhan}, N., {Allard}, N.~F., {et~al.} 2019, \apjl, 887,
  L20

\bibitem[{{Yu} {et~al.}(2021){Yu}, {Moses}, {Fortney}, \& {Zhang}}]{Yu2021}
{Yu}, X., {Moses}, J.~I., {Fortney}, J.~J., \& {Zhang}, X. 2021, \apj, 914, 38

\bibitem[{{Yurchenko} {et~al.}(2011){Yurchenko}, {Barber}, \&
  {Tennyson}}]{yurchenko2011}
{Yurchenko}, S.~N., {Barber}, R.~J., \& {Tennyson}, J. 2011, \mnras, 413, 1828

\bibitem[{{Yurchenko} \& {Tennyson}(2014)}]{yurchenko2014}
{Yurchenko}, S.~N., \& {Tennyson}, J. 2014, \mnras, 440, 1649

\bibitem[{{Zhou} {et~al.}(2017){Zhou}, {Apai}, {Lew}, \&
  {Schneider}}]{2017AJ....153..243Z}
{Zhou}, Y., {Apai}, D., {Lew}, B. W.~P., \& {Schneider}, G. 2017, \aj, 153, 243

\end{thebibliography}

\acknowledgements 

The authors are grateful for constructive feedback provided by the anonymous referee. Support for HST program GO-15814 was provided by NASA through a grant from the Space Telescope Science Institute, which is operated by the Association of Universities for Research in Astronomy, Inc., under NASA contract NAS 5-26555. This research has made use of the SIMBAD database, operated at CDS, Strasbourg, France. Some of the data presented in this paper were obtained from the Mikulski Archive for Space Telescopes (MAST). STScI is operated by the Association of Universities for Research in Astronomy, Inc., under NASA contract NAS5-26555. Support for MAST for non-HST data is provided by the NASA Office of Space Science via grant NNX13AC07G and by other grants and contracts. MNG acknowledges support from the European Space Agency (ESA) as an ESA Research Fellow. NM thanks Savvas Constantinou for helpful discussions.

\software{NumPy \citep{numpy2011}, SciPy \citep{2020SciPy-NMeth}, Matplotlib \citep{2007CSE.....9...90H}, emcee \citep{2013PASP..125..306F}, batman \citep{2015PASP..127.1161K}, Astropy \citep{2013A&A...558A..33A,2018AJ....156..123A}, pysynphot \citep{2013ascl.soft03023S}, PyMultinest \citep{2014A&A...564A.125B}}

\facilities{ HST(WFC3), Spitzer(IRAC), TESS }

All the {\it HST} data used in this paper can be found in MAST: \dataset[10.17909/2h3r-t275]{http://dx.doi.org/10.17909/2h3r-t275}

\appendix

\section{Stellar limb darkening} \label{app:limbDarkening}

\begin{figure}
\centering  
\includegraphics[width=0.7\linewidth]{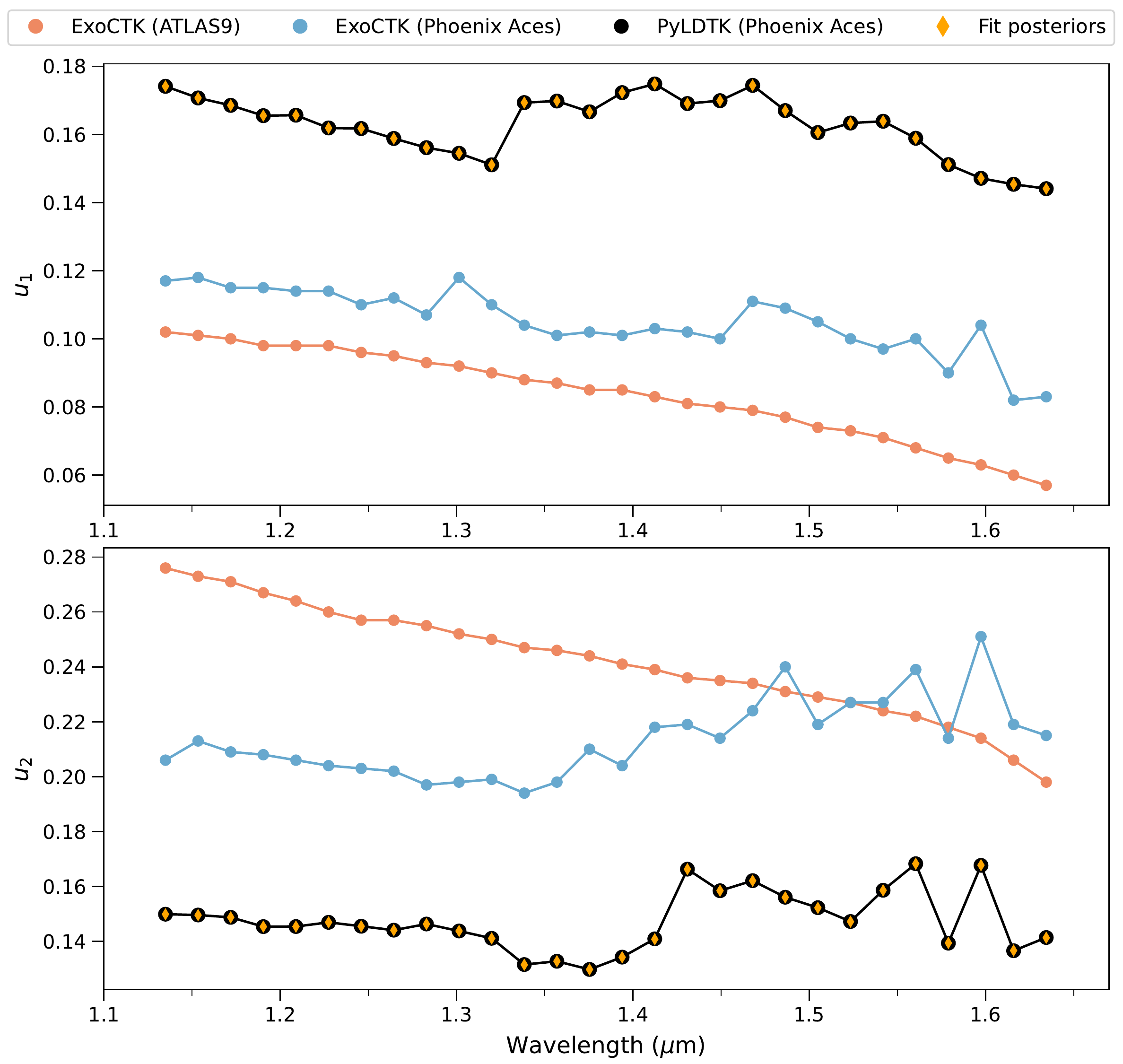}
\caption{Coefficients ($u_1$, $u_2$) for a quadratic stellar limb darkening law obtained using ExoCTK and PyLDTK. Separate sets of ExoCTK coefficients were computed assuming an ATLAS 9 (blue circles) and Phoenix ACES (red circles) stellar model. The PyLDTK coefficients (black points) assume a Phoenix ACES stellar model. Posterior distributions obtained from the light curve fits are shown as orange diamonds, with $1\sigma$ credible intervals that are smaller than the marker symbols.}
\label{fig:limbDarkeningCoeffs}
\end{figure}

For the fiducial light curve analyses, a quadratic stellar limb darkening law was adopted with coefficients that were allowed to vary as free parameters. As described in Sections \ref{sec:lcBroad} and \ref{sec:lcSpec}, tight Gaussian priors obtained using PyLDTK were adopted for these light curve fits. The limb darkening coefficiet priors and resulting posteriors are listed in Table \ref{table:wfc3lcSpecLD} for the spectroscopic light curves. Due to the limited phase coverage of the transit, the limb darkening coefficients are effectively unconstrained by the data and consequently the posteriors are indistinguishable from the priors (Figure \ref{fig:limbDarkeningCoeffs}).

\begin{figure}
\centering  
\includegraphics[width=\linewidth]{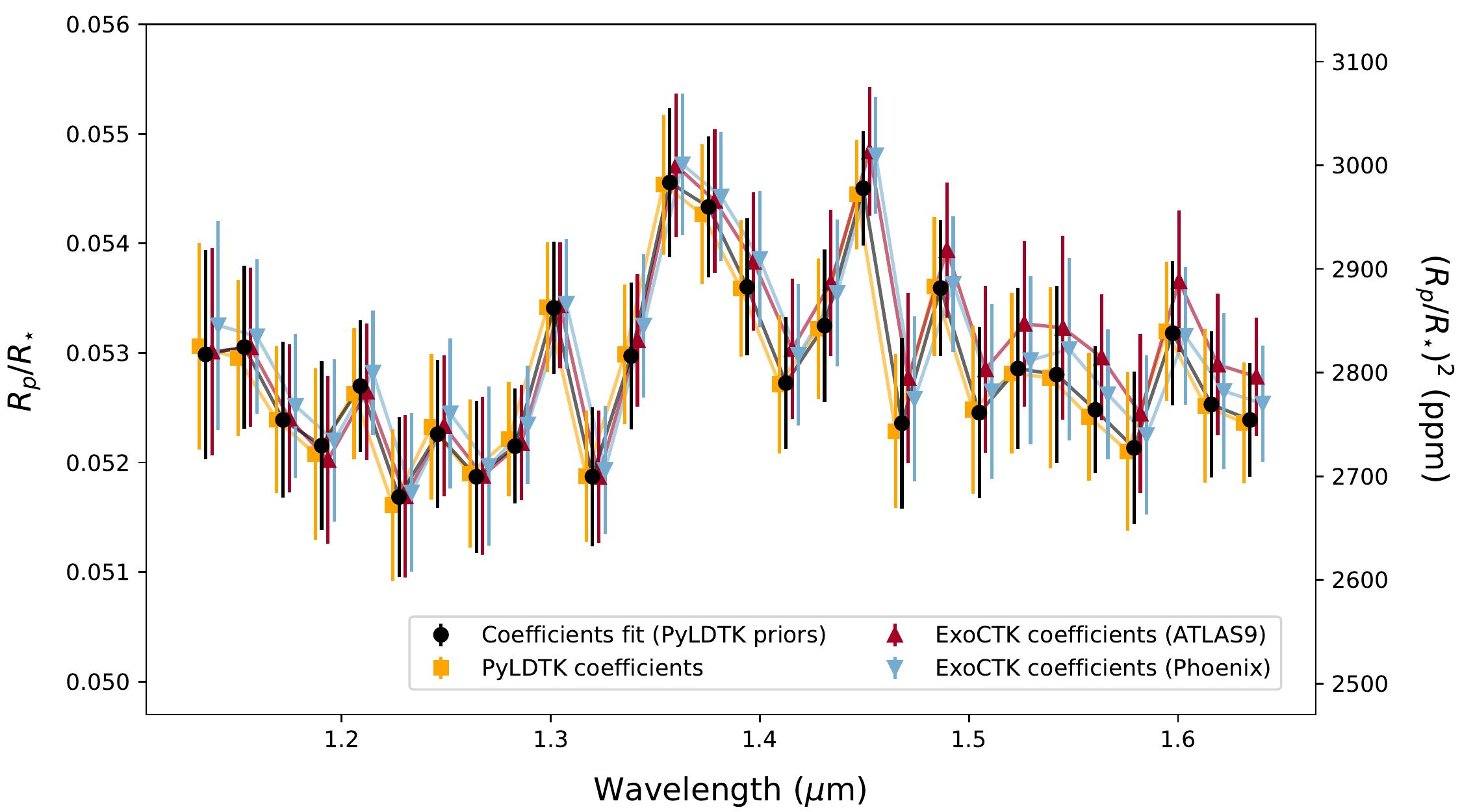}
\caption{Derived transmission spectra for the different stellar limb darkening treatments described in Appendix \ref{app:limbDarkening}. Free coefficients with PyLDTK priors (black circles), coefficients held fixed to the PyLDTK values (orange squares), and coefficients held fixed to the ExoCTK values assuming ATLAS9 (red triangles) and Phoenix ACES (blue triangles) stellar models. A high level of consistency is obtained for all limb darkening treatments.}
\label{fig:limbDarkeningTrSpec}
\end{figure}

To investigate the sensitivity of the final results to the stellar limb darkening treatment, the spectroscopic light curve fits were repeated with $u_1$ and $u_2$ held fixed to values computed using the online ExoCTK tool.\footnote{https://exoctk.stsci.edu} Specifically, ExoCTK provides the option of computing limb darkening coefficients using both Phoenix ACES \citep{Husser2013} and ATLAS9 \citep{1993KurCD..13.....K,2003IAUS..210P.A20C} stellar models. As shown in Figure \ref{fig:limbDarkeningCoeffs}, there are clear differences in the limb darkening coefficients computed by ExoCTK using these two stellar models, and compared to those obtained with PyLDTK, which also uses the Phoenix ACES models. However, as can be seen in Figure \ref{fig:limbDarkeningTrSpec}, the  final transmission spectrum is found to be robust to the choice of limb darkening treatment. 

\begin{table}
\begin{minipage}{\columnwidth}
\centering
\caption{ Posterior distributions for stellar limb darkening coefficients inferred from the  spectroscopic light curve fits. \label{table:wfc3lcSpecLD} }
\begin{tabular}{ccc}
\hline Wavelength & $u_1$ & $u_2$ \\ 
($\mu$m) &  &  \\ \hline 
1.126-1.144 & $0.17411_{-0.00015}^{+0.00017}$ & $0.14986_{-0.00049}^{+0.00049}$ \\  
1.144-1.163 & $0.17069_{-0.00017}^{+0.00016}$ & $0.14954_{-0.00055}^{+0.00058}$ \\  
1.163-1.181 & $0.16853_{-0.00017}^{+0.00016}$ & $0.14873_{-0.00053}^{+0.00050}$ \\  
1.181-1.200 & $0.16550_{-0.00016}^{+0.00016}$ & $0.14536_{-0.00053}^{+0.00051}$ \\  
1.200-1.218 & $0.16563_{-0.00016}^{+0.00016}$ & $0.14539_{-0.00053}^{+0.00051}$ \\  
1.218-1.237 & $0.16187_{-0.00017}^{+0.00017}$ & $0.14692_{-0.00053}^{+0.00052}$ \\  
1.237-1.255 & $0.16173_{-0.00017}^{+0.00017}$ & $0.14545_{-0.00055}^{+0.00052}$ \\  
1.255-1.274 & $0.15879_{-0.00018}^{+0.00019}$ & $0.14405_{-0.00060}^{+0.00062}$ \\  
1.274-1.292 & $0.15610_{-0.00020}^{+0.00020}$ & $0.14629_{-0.00063}^{+0.00062}$ \\  
1.292-1.311 & $0.15446_{-0.00019}^{+0.00017}$ & $0.14377_{-0.00059}^{+0.00059}$ \\  
1.311-1.329 & $0.15106_{-0.00019}^{+0.00017}$ & $0.14106_{-0.00055}^{+0.00056}$ \\  
1.329-1.348 & $0.16933_{-0.00017}^{+0.00017}$ & $0.13147_{-0.00072}^{+0.00068}$ \\  
1.348-1.366 & $0.16980_{-0.00021}^{+0.00020}$ & $0.13276_{-0.00084}^{+0.00082}$ \\  
1.366-1.385 & $0.16661_{-0.00019}^{+0.00020}$ & $0.12976_{-0.00090}^{+0.00077}$ \\  
1.385-1.403 & $0.17225_{-0.00021}^{+0.00022}$ & $0.13419_{-0.00093}^{+0.00094}$ \\  
1.403-1.422 & $0.17482_{-0.00027}^{+0.00027}$ & $0.14076_{-0.00107}^{+0.00115}$ \\  
1.422-1.440 & $0.16906_{-0.00028}^{+0.00029}$ & $0.16630_{-0.00120}^{+0.00123}$ \\  
1.440-1.459 & $0.16990_{-0.00028}^{+0.00030}$ & $0.15845_{-0.00131}^{+0.00129}$ \\  
1.459-1.477 & $0.17439_{-0.00032}^{+0.00030}$ & $0.16207_{-0.00121}^{+0.00125}$ \\  
1.477-1.496 & $0.16697_{-0.00027}^{+0.00031}$ & $0.15601_{-0.00115}^{+0.00116}$ \\  
1.496-1.514 & $0.16053_{-0.00030}^{+0.00028}$ & $0.15220_{-0.00126}^{+0.00130}$ \\  
1.514-1.533 & $0.16335_{-0.00029}^{+0.00026}$ & $0.14715_{-0.00110}^{+0.00113}$ \\  
1.533-1.551 & $0.16388_{-0.00029}^{+0.00028}$ & $0.15854_{-0.00119}^{+0.00117}$ \\  
1.551-1.570 & $0.15884_{-0.00027}^{+0.00030}$ & $0.16827_{-0.00108}^{+0.00115}$ \\  
1.570-1.588 & $0.15116_{-0.00028}^{+0.00024}$ & $0.13940_{-0.00111}^{+0.00114}$ \\  
1.588-1.607 & $0.14712_{-0.00027}^{+0.00028}$ & $0.16774_{-0.00113}^{+0.00111}$ \\  
1.607-1.625 & $0.14538_{-0.00019}^{+0.00018}$ & $0.13665_{-0.00086}^{+0.00077}$ \\  
1.625-1.644 & $0.14406_{-0.00023}^{+0.00024}$ & $0.14131_{-0.00090}^{+0.00097}$ \\ \hline 
\end{tabular}
\end{minipage}
\end{table}

\begin{deluxetable*}{c|c|c|c|c}
\tablecaption{Priors and Retrieved Parameters for the Models in this Work
\label{table:priors}}
\tablecolumns{5}
\tablewidth{\columnwidth}
\tablehead{
\colhead{Parameter} & \colhead{Prior Distribution} & \colhead{Baseline Model$^{\rm T, W}$} & \colhead{Stellar Heterogeneity$^{\rm T, W}$} & \colhead{Stellar Heterogeneity$^{\rm W}$}
}
\startdata
$\log_{10}(X_{\text{H}_2\text{O}})$ & $\mathcal{U}(-12,-0.3)$ & $-1.77^{+0.69}_{-0.93}$ &   $-1.91^{+0.74}_{-0.98}$ &  $-2.09^{+0.89}_{-1.21}$\\ \hline
$\log_{10}(X_{\text{CH}_4})$ & $\mathcal{U}(-12,-0.3)$ & $-7.78^{+2.67}_{-2.49}$ &   $-7.71^{+2.60}_{-2.57}$ &  $-7.35^{+2.54}_{-2.57}$\\ \hline
$\log_{10}(X_{\text{NH}_3})$ & $\mathcal{U}(-12,-0.3)$ & $-8.04^{+2.42}_{-2.37}$ &   $-7.72^{+2.25}_{-2.41}$ &  $-8.02^{+2.25}_{-2.29}$\\ \hline
$\log_{10}(X_{\text{CO}})$ & $\mathcal{U}(-12,-0.3)$ & $-6.60^{+3.37}_{-3.31}$ &   $-6.15^{+3.10}_{-3.33}$ &  $-6.36^{+3.13}_{-3.28}$\\ \hline
$\log_{10}(X_{\text{CO}_2})$ & $\mathcal{U}(-12,-0.3)$ & $-5.81^{+3.22}_{-3.79}$ &   $-5.62^{+3.11}_{-3.77}$ &  $-5.73^{+3.04}_{-3.67}$\\ \hline
$\log_{10}(X_{\text{HCN}})$ & $\mathcal{U}(-12,-0.3)$ & $-6.53^{+2.93}_{-3.27}$ &   $-6.58^{+2.87}_{-3.16}$ &  $-6.70^{+2.82}_{-3.06}$\\ \hline
$\text{T}_0$ [K]& $\mathcal{U}(0,400)$ &  $247^{+80}_{-68}$  & $257^{+72}_{-69}$  & $232^{+76}_{-65}$  \\ \hline
$\log_{10}(\text{P}_{\rm ref})$[bar] & $\mathcal{U}(-6,2)$ &$-1.04^{+0.46}_{-0.54}$ & $-2.35^{+1.47}_{-1.78}$ & $-3.34^{+1.56}_{-1.47}$  \\ \hline
$\log_{10}(a)$ & $\mathcal{U}(-4,10)$ & $0.71^{+4.04}_{-2.90}$ & $0.49^{+3.39}_{-2.70}$   & $1.36^{+3.95}_{-3.13}$ \\ \hline
$\gamma$ & $\mathcal{U}(-20,2)$ & $-10.83^{+7.20}_{-5.80}$ & $-10.84^{+6.99}_{-5.59}$ & $-10.52^{+6.97}_{-5.75}$ \\ \hline
$\log_{10}(\text{P}_{\rm c})$ [bar]& $\mathcal{U}(-6,2)$ & $-0.77^{+1.63}_{-2.47}$  & $-0.88^{+1.68}_{-2.45}$  & $-1.53^{+1.91}_{-2.22}$  \\ \hline
$\phi$ & $\mathcal{U}(0,1)$ & $0.34^{+0.36}_{-0.22}$ & $0.38^{+0.33}_{-0.23}$  & $0.37^{+0.31}_{-0.23}$ \\ \hline
$\delta$ & $\mathcal{U}(0.0,0.5)$ & N/A & $0.06^{+0.05}_{-0.03}$  & $0.08^{+0.05}_{-0.04}$ \\ \hline
$\text{T}_{\rm het}$ [K] & $\mathcal{U}(701,5259)$ &  N/A  & $2446^{+1120}_{-988}$  & $2202^{+855}_{-866}$  \\ \hline
$\text{T}_{\rm phot}$ [K]& $\mathcal{N}(3506,100)$ &  N/A  & $3512^{+81}_{-84}$   &  $3505^{+84}_{-82}$  
\enddata 
\tablecomments{The superscript in the model name indicates the data included in the retrieval: T for TESS and W for HST-WFC3. N/A means that the parameter was not considered in the model by construction.}
\end{deluxetable*}

\end{document}